\documentclass[12pt]{article}  
\usepackage{color,amssymb,amsthm,amsmath,amsxtra,amsfonts,breqn,
fullpage,graphicx,hyperref,float,appendix,mathtools,eqparbox, 
lscape,pdfpages}

\usepackage[displaymath,mathlines]{lineno}

\usepackage{times}

\usepackage[latin1]{inputenc}
\usepackage{tikz}
\usetikzlibrary{shapes,arrows}

\newcommand{\overbar}[1]{\mkern 1.5mu\overline{\mkern-1.5mu#1\mkern-1.5mu}\mkern 1.5mu}

\usepackage{fancyhdr}
\usepackage{xcolor}
\definecolor{dark-red}{rgb}{0.4,0.15,0.15}
\definecolor{dark-blue}{rgb}{0,0,0.45}
\hypersetup{
    colorlinks, linkcolor={dark-blue},
    citecolor={blue}, urlcolor={blue}}

\numberwithin{equation}{section}
\usepackage{pdflscape}

\definecolor{gold}{rgb}{0.85,0.66,0.0}

\begin{document}

\title{A Multi-Spatial, Multi-Temporal, Semi-Analytical Model for Bathymetry, Water Turbidity and Bottom Composition using Multispectral Imagery}
\author{Sam Blake\\ The University of Melbourne}
\date{December, 2019}
\maketitle

\begin{abstract}
In this paper we introduce a semi-analytical model for bathymetry, water turbidity and bottom 
composition; which is primarily based on the physics-based model, HOPE, of Lee et al 
\cite{lee1998}\cite{lee1999}. Unlike the model of Lee, which was originally designed to 
use hyperspectral imagery, our model is specifically designed to use multispectral satellite 
imagery. In particular, we adapt to the greatly decreased spectral resolution by introducing 
temporal and spatial assumptions on the depth and water turbidity. We validate the extensions 
to the Lee et al model with a 260 $\text{km}^2$ case study in the area of the Murion Islands  
off Western Australia, where we compare the atmospherically-corrected LANDSAT-8 derived 
bathymetry against a 2011 single-beam sonar survey by Transport Western Australia. The model 
validates well against the single-beam sonar survey, with $R^2=0.77$, a mean absolute error 
of 1.37 m and a mean relative error of 9.24\%. This indicates the model could be widely 
applicable to LANDSAT-8 imagery. 
\end{abstract}


\begin{table}[H]
\centering
\caption{Symbols}
{\small
\begin{tabular}{ l l l }
 symbol & description & units \\
 \hline
 $R_{\text{rs}}$ & remote sensing reflectance (RRS) & $\text{sr}^{-1}$ \\
 $r_{\text{rs}}$ & subsurface remote sensing reflectance & $\text{sr}^{-1}$ \\  
 $P$ & phytoplankton absorption coefficient at 440 nm & $\text{m}^{-1}$ \\
 $G$ & absorption coefficient of gelbstoff and detritus at 440 nm & $\text{m}^{-1}$ \\
 $X$ & backscattering coefficient of particles at 440 nm & $\text{m}^{-1}$ \\
 $B$ & bottom albedo at 550 nm & \\
 $H$ & bottom depth & m \\
 $\Delta$ & spectrally-constant offset & $\text{sr}^{-1}$ \\
 $a$ & absorption coefficient & $\text{m}^{-1}$ \\
 $b_b$ & backscattering coefficient & $\text{m}^{-1}$ \\
 $k$ & attenuation coefficient & $\text{m}^{-1}$ \\
 $\rho$ & bottom albedo spectrum normalised to 550 nm & \\
 $\theta_{\text{sun}}$ & subsurface solar zenith angle & rad \\
 $\theta_{\text{view}}$ & subsurface viewing angle from nadir & rad \\
 \hline
\end{tabular}
}
\end{table}

\section{Introduction}

The origins of marine remote sensing by satellite date to the early 1970s with the launch of LANDSAT 1 
and the seminal research of Polcyn \& Cousteau \cite{polcyn1970}\cite{polycn1976}. In 1975, NASA and Jacques 
Cousteau used multispectral imagery from LANDSAT 1 and in situ measurements of water clarity and 
sea floor reflectance to estimate the bathymetry in the Bahamas and off the eastern coast of Florida. They 
concluded that with optimal conditions, satellite-derived depths could be reliably modelled up to a depth of 22 m. 
Polcyn initially used a single band to estimate the depth, where the band attenuation coefficient and 
bottom reflectance was measured at the same time as the satellite image acquisition. Later Polcyn generalised 
this method to a depth estimate based on the ratio of two bands. 
As the field measurements were taken at a single location, these depth estimates assumed constant 
water turbidity and bottom reflectance across the entire satellite scene. \\

In 1978, Lyzenga \cite{lyzenga1978} introduced an empirical multi-band method to estimate the 
bathymetry. Like the ratio method of Polcyn, this method assumed constant water turbidity and 
bottom type. The Lyzenga method is outlined in Section \ref{lyzenga_method}. In 2006, Lyzenga 
et al refined this method \cite{lyzenga2006}. \\

Since the turn of the century, physics-based radiative transfer models have existed to 
simultaneously estimate depth, water turbidity and bottom reflectance from hyperspectral 
imagery. A detailed summary and inter-comparison at two sites of a number of key models 
is given in Dekker et al \cite{dekker2011}. \\

The original model of Lee et al \cite{lee1998}\cite{lee1999}, HOPE (Hyperspectral Optimisation 
Process Exemplar), assumed a single benthic substratum (sand). Lee et al \cite{lee2001} 
subsequently generalised this to two benthic substratum - sand and seagrass, where the two 
were delineated by way of an empirical relationship prior to the model inversion. The HOPE 
model originally used the SOLVER tool (within Microsoft Excel) as the optimisation routine. 
Subsequently, HOPE used the Levenberg-Marquardt and B2NLS optimisation 
algorithms \cite{lee2009}.  \\

In 2004, Klonowski et al \cite{klonowski2004}\cite{klonowski2007} made extensions to the model of Lee et al 
to classify bottom spectra in Jurien Bay, Western Australia. Modifications were made to the model of 
Chlorophyll-a concentrations, based on measurements in Cockburn Sound. This model included a 
parameterisation for three bottom spectra - sand, Hallophylla (green leafy seagrass) and Sargassum 
(brown seaweed). This model used the Levenburg-Marquardt optimisation scheme. \\ 

In 2005, Mobley et al \cite{mobley2005} used the radiative transfer numerical model, HydroLight, 
to construct lookup tables for spectrum matching of hyperspectral imagery. This model is 
known as CRISTAL (Comprehensive Reflectance Inversion based on Spectrum matching and TAble 
Lookup) \cite{dekker2011}. \\

In 2006, Wettle et al \cite{wettle2006}\cite{brando2009} developed a semi-analytical model, SAMBUCA (Semi-Analytical 
Model for Bathymetry, Un-mixing, and Concentration Assessment), based on the Lee et al model, HOPE. 
This model incorporates a different parameterisation for the absorption and backscatter 
coefficients to HOPE, which is in line with field data measurement protocols used by the CSIRO. 
This model used the Simplex optimisation scheme. \\

In 2009, Hedley et al \cite{hedley2009} implemented an efficient lookup table inversion scheme 
for radiative transfer models using adaptive lookup trees (ALUT). This model was more efficient 
than optimisation-based models and had comparable accuracy in bathymetric retrievals. \\ 

In 2014, Gege \cite{gege2004} used the shallow water parameterisation of the RRS by Albert and 
Mobley \cite{albert2003} to implement the WASI (Water Colour Simulator) model. This parameterisation 
is an alternative to the model of Lee et al. This model uses the simplex optimisation scheme. \\

In 2018, Hedley et al \cite{hedley2018} examined the use of the Sentinel-2 satellite to monitor 
coral reefs. It was found that the Sentinel-2 data can be used within a physics-based model to 
monitor coral reefs and retrieve bathymetric estimations with comparable performance to WorldView-2. \\

In this paper we detail the development of the \textit{photic} model, an extension of 
the Lee et al \cite{lee1998}\cite{lee1999}\cite{lee2001} semi-analytical model for use with 
multispectral imagery. In particular, for the LANDSAT-8 and Sentinel-2 satellites. These 
satellites have greatly decreased spectral resolution in comparison to airborne 
hyperspectral sensors, however they have regular global coverage in coastal areas and their 
datasets are freely available. Given this availability, we are motivated to see if we can 
apply semi-analytical models accurately and efficiently to these datasets.

In section 2 we recap the hyperspectral model of Lee et al \cite{lee1998}\cite{lee1999}\cite{lee2001}, 
along with extensions to include multiple bottom types by Gege \cite{gege2004} and 
Wettle et al \cite{wettle2006}. We follow Wettle et al \cite{wettle2006} and use 
both the root-mean-square and spectral angle mapper error in our error metric.  \\

In section 3 we introduce spatial and temporal extensions to the Lee et al model. \\ 

In section 4 we detail how the model is inverted to yield stable depth, water turbidity and 
bottom reflectance estimates. \\

In section 5 we detail an iterative modelling framework, where model parameters are 
estimated and subsequently refined to improve their retrieval. \\

In section 6 we give a detailed case study where we validate the \textit{photic} model 
against single-beam sonar data in the area of the Murion Island Marine Management Area. 

\section{The Model for Hyperspectral Imagery}

Our modelling approach for hyperspectral imagery directly follows the model of Lee et 
al \cite{lee1998}\cite{lee1999}\cite{lee2001}, with additions to include multiple bottom 
spectra as given in the model of Gege \cite{gege2004} and the spectral angle mapper 
(SAM) in the spectral matching error metric as given in the model of Wettel et 
al \cite{wettle2006}. We will now briefly describe the model. \\

The remote-sensing reflectance (RRS), $R_{rs}$, is defined at the ratio of the 
water leaving radiance to downwelling irradiance just above the surface. 
The RRS over optically shallow 
water is controlled by a number of factors, including absorption properties, scattering 
properties, the bottom albedo, bottom depth and solar elevation. We begin by relating the RRS 
above the surface, $R_{rs}(\lambda)$, to the subsurface RRS, $r_{rs}(\lambda)$ which is given by
\begin{linenomath}
\begin{equation*}
R_{rs}(\lambda) = \frac{0.5\,r_{rs}(\lambda)}{1 - 1.5\,r_{rs}(\lambda)} + \Delta,
\end{equation*} 
\end{linenomath}
where $\Delta$ is a spectrally-constant offset. The subsurface RRS, $r_{rs}(\lambda)$, is 
expressed as a linear combination of subsurface RRS from the water column, $r_{rs_C}(\lambda)$, 
and the subsurface RRS from the bottom reflectance, $r_{rs_B(\lambda)}$. 
\begin{linenomath}
\begin{align}
r_{rs}(\lambda) &= r_{rs_C}(\lambda) + r_{rs_B}(\lambda)\nonumber \\
&= r_{rs_{\infty}}(\lambda) \left(1 - \exp\left(-\left(\frac{1}{\cos(\theta_{\text{sun}})} + \frac{D_C(\lambda)}{\cos(\theta_{\text{view}})}\right)\,k(\lambda)\,H\right)\right) + \nonumber\\
& \qquad \frac{\rho(\lambda)}{\pi}\exp\left(-\left(\frac{1}{\cos(\theta_{\text{sun}})} + \frac{D_B(\lambda)}{\cos(\theta_{\text{view}})}\right)\,k(\lambda)\,H\right) \label{rrs_model}
\end{align}
\end{linenomath}
Where $r_{rs_{\infty}}(\lambda)$ is the subsurface RRS for optically deep water. $D_C(\lambda)$ and $D_B(\lambda)$ 
are the path elongation for photons from the water column and from the bottom respectively. 
$\theta_{\text{sun}}$ is the subsurface solar zenith angle and $\theta_{\text{view}}$ is 
the subsurface viewing angle from nadir. $\rho(\lambda)$ is the bottom albedo. $H$ is the bottom 
depth. $k(\lambda)$ is the attenuation coefficient, which is given by
\begin{linenomath}
\begin{equation}
k(\lambda) = a(\lambda) + b_b(\lambda)\label{k_eqn}
\end{equation}
\end{linenomath}
where $a(\lambda)$ is the absorption coefficient and $b_b(\lambda)$ is the backscattering 
coefficient. \\

For the subsurface RRS of optically deep water, Lee et al \cite{lee1998}\cite{lee1999}\cite{lee2001} 
used the model of Gordon et al \cite{gordon1988} 
\begin{linenomath}
\begin{equation*}
r_{\infty}(\lambda) = 0.084\,u(\lambda) + 0.170\,u(\lambda)^2,
\end{equation*} 
\end{linenomath}
where 
\begin{linenomath}
\begin{equation*}
u(\lambda) = \frac{b_b(\lambda)}{a(\lambda) + b_b(\lambda)}.
\end{equation*} 
\end{linenomath}

The path elongation factors are given by 
\begin{linenomath}
\begin{align*}
D_C(\lambda) &= 1.03 \sqrt{1 + 2.4 u}\\
D_B(\lambda) &= 1.04 \sqrt{1 + 5.4 u}.
\end{align*}
\end{linenomath}

The inherent optical properties, $a(\lambda)$ \& $b_b(\lambda)$ are given by 
\begin{linenomath}
\begin{equation*}
a(\lambda) = a_w(\lambda) + a_{\phi}(\lambda) + a_g(\lambda)
\end{equation*} 
\end{linenomath}
\begin{linenomath}
\begin{equation*}
b(\lambda) = b_{bw}(\lambda) + b_{bp}(\lambda),
\end{equation*} 
\end{linenomath}
where $a_w(\lambda)$ is the absorption coefficients of pure water as given in Pope \& 
Fry \cite{pope1997}, $a_{\phi}(\lambda)$ is the absorption coefficients for phytoplankton 
pigments, $a_g(\lambda)$ is the absorption coefficients for gelbstoff and detrius, 
$b_{bw}(\lambda)$ is the backscattering coefficient for pure seawater as given in 
Morel \cite{morel1974}, $b_{bp}(\lambda)$ is the backscattering coefficient of 
suspended particles. \\

For an $n$-band spectrum the model above has $n$ equations (one for each $\lambda$), each 
with 4 unknowns -- $a_{\phi}(\lambda), a_g(\lambda), b_{bp}(\lambda), \rho(\lambda), H$. 
Thus, there are $4n + 1$ unknowns and $n$ equations, and consequently finding solutions 
to this system requires establishing additional relationships. \\

Lee et al estimated the spectral shape of $a_{\phi}(\lambda)$ with a single parameter, $P$, which 
represents the phytoplankton absorption coefficient at 440 nm, ie. $P = a_{\phi}(440)$ 
\begin{linenomath}
\begin{equation}
a_{\phi}(P,\lambda) = \left(a_0(\lambda) + a_1(\lambda)\log(P)\right)P,\label{a_phi}
\end{equation}
\end{linenomath}
where $a_0(\lambda)$, $a_1(\lambda)$ was modelled by Lee \cite{lee1994}. We use this model for $a_{\phi}$, 
however we derived the parameterisation of $a_0(\lambda)$, $a_1(\lambda)$ from a dataset sourced 
from the CSIRO \cite{csiro2005} (See Appendix A). \\

The spectral shape for the absorption of gelbstoff and detritus, $a_g(\lambda)$ is 
expressed with the parameter $G = a_g(440)$ and is given by
\begin{linenomath}
\begin{equation*}
a_g(G,\lambda) = G\, e^{-S(\lambda - 440)}.
\end{equation*}
\end{linenomath}

The parameter S is in the range $0.011$--$0.021$ $\text{nm}^{-1}$. \\

The spectral shape for the backscattering coefficient of suspended particles, 
$b_{bp}(\lambda)$ is expressed with the parameter $X = b_{bp}(440)$ and is given by
\begin{linenomath}
\begin{equation*}
b_{bp}(X,\lambda) = X\, \left(\frac{440}{\lambda}\right)^Y,
\end{equation*}
\end{linenomath}
where $Y$ is estimated by the empirical relationship 
\begin{linenomath}
\begin{equation*}
Y = 3.44 \left( 1.0 - 3.17 e^{-2.01 \chi} \right) \quad \text{and} \quad \chi = \frac{R_{rs}(440) - R_{rs}(750)}{R_{rs}(490) - R_{rs}(750)}.
\end{equation*}
\end{linenomath}

Finally, for $N_b$ bottom types, the bottom albedo is parameterised with 
$B = \rho(550)$ and we write $\rho(\lambda)$ as $\rho(B, q_0, q_1, \cdots, q_{N_b-1}, \lambda)$, where 
\begin{linenomath}
\begin{equation}
\rho(B_0, B_1, \cdots, B_{N_b-1}, q_0, q_1, \cdots, q_{N_b-1}, \lambda) = \frac{\sum\limits_{i=0}^{N_b-1}B_i\,q_i\,\rho_{\text{bottom}_i}(\lambda)}{\sum\limits_{i=0}^{N_b-1}q_i},\label{bottom_formulation}
\end{equation}
\end{linenomath}
where $\rho_{\text{bottom}_i}(\lambda)$ is the bottom albedo of a given bottom type as 
estimated from field measurements and normalised to 550 nm, and the $q_i$ specify the fraction 
of each bottom type. This is based on the parameterisations used in Gege \cite{gege2004} and 
Wettle et al \cite{wettle2006}. \\

Our model can estimate bottom reflectance using (\ref{bottom_formulation}), however if the 
model is required to delineate between a large number of bottom spectra then we are better to 
adopt the iterative method of Wettle et al \cite{wettle2006}. In this formulation, we iterate over 
all individual bottom spectra, then over all pairs of bottom spectra. The spectra or pair of
spectra best matching the input RRS is then representative of the bottom composition. It is worth 
noting that this method is significantly slower due to the combinatorial nature of exhaustively 
iterating through all the pairs. \\

With this parameterisation we reduce the number of unknowns which influence the RRS spectrum to $P$, 
$G$, $X$, $B_0$, $B_1$, $\cdots$, $B_{N_b-1}$, $q_0, q_1, \cdots, q_{N_b-1}$, $H$ and $\Delta$. Thus if we have 
$5 + 2 N_{b}$ independent channels it should be possible to determine a unique solution to the 
system of equations. However due to the presence of noise the best we can do is seek to minimise 
the difference between the modelled RRS, $R_{rs_{\text{modelled}}}(\lambda)$, and the measured 
RRS, $R_{rs_{\text{measured}}}(\lambda)$. Following Lee et al \cite{lee1998}\cite{lee1999}\cite{lee2001}, 
we define the RRS RMS relative error, $E_{R_{rs}}^{\text{RMS}}$, as 
\begin{linenomath}
\begin{multline}
E_{R_{rs}}^{\text{RMS}}(P, G, X, B_0, B_1, \cdots, B_{N_b-1}, q_0, q_1, \cdots, q_{N_b-1}, H, \Delta) = \\
	\frac{\sqrt{\sum\limits_{\lambda} \left(R_{rs_{\text{modelled}}}(\lambda) - R_{rs_{\text{measured}}}(\lambda)\right)^2}}
	{\sum\limits_{\lambda} R_{rs_{\text{measured}}}(\lambda)}.
\end{multline}
\end{linenomath}

\noindent Wettle et al \cite{wettle2006} included the Spectral Angle Mapper (SAM) 
within the error metric
\begin{linenomath}
\begin{multline*}
E_{R_{rs}}^{\text{SAM}}(P, G, X, B_0, B_1, \cdots, B_{N_b-1}, q_0, q_1, \cdots, q_{N_b-1}, H, \Delta) = \\
\cos^{-1}\left( \frac{R_{rs_{\text{modelled}}} \cdot R_{rs_{\text{measured}}}}
	{\left(R_{rs_{\text{modelled}}} \cdot R_{rs_{\text{modelled}}}\right) \left(R_{rs_{\text{measured}}} \cdot R_{rs_{\text{measured}}}\right)} \right),
\end{multline*}
\end{linenomath}
\noindent where in both $E_{R_{rs}}^{\text{RMS}}$ and $E_{R_{rs}}^{\text{SAM}}$ the 
term $R_{rs_{\text{modelled}}}$ is dependent on the parameters $P$, $G$, $X$, $B_0$, 
$B_1$, $\cdots$, $B_{N_b-1}$, $q_0$, $q_1$, $\cdots$, $q_{N_b-1}$, $H$, $\Delta$ and 
$\lambda$. We will describe the full error metric used in \textit{photic} in the 
following section.

\section{Extensions to the Model for Multispectral Imagery}

Lee et al \cite{lee2002} found that 15 equispaced spectral bands covering the 
400--800 nm range are adequate for most coastal and oceanic remote sensing 
applications. The satellite imagery of LANDSAT 8 and Sentinel-2 have significantly 
fewer spectral bands in this range, however we would still like to apply the 
semi-analytical model to these vast and freely available datasets. Previously 
Dekker et al \cite{dekker2012} has applied the SAMBUCA \cite{wettle2006} model to 
QuickBird multispectral satellite imagery. \\

We now describe two extensions to the semi-analytical model to increase its applicability to 
multispectral imagery. 

	\subsection{Spatial extension}

A reasonable assumption to make for medium to high resolution satellite imagery is constant 
water turbidity within a spatial region around a given pixel. Klonowski et al \cite{klonowski2004} 
used a constant $P$, $G$ and $X$ throughout an entire scene for bottom type classification 
using hyperspectral imagery. \\

If we define a spatial region, $r$, around a given pixel, then we (simultaneously) model 
$N_r = (2r+1)^2$ pixels. Within this region we constrain $P$, $G$, $X$ and $\Delta$ to 
be (spatially) static, while $H$, $B_0$, $B_1$, $\cdots$, $B_{N_b-1}$, $q_0$, $q_1$, 
$\cdots$, $q_{N_b-1}$ may vary. 

\begin{figure}[H]
\centering
\includegraphics[width=0.6\textwidth]{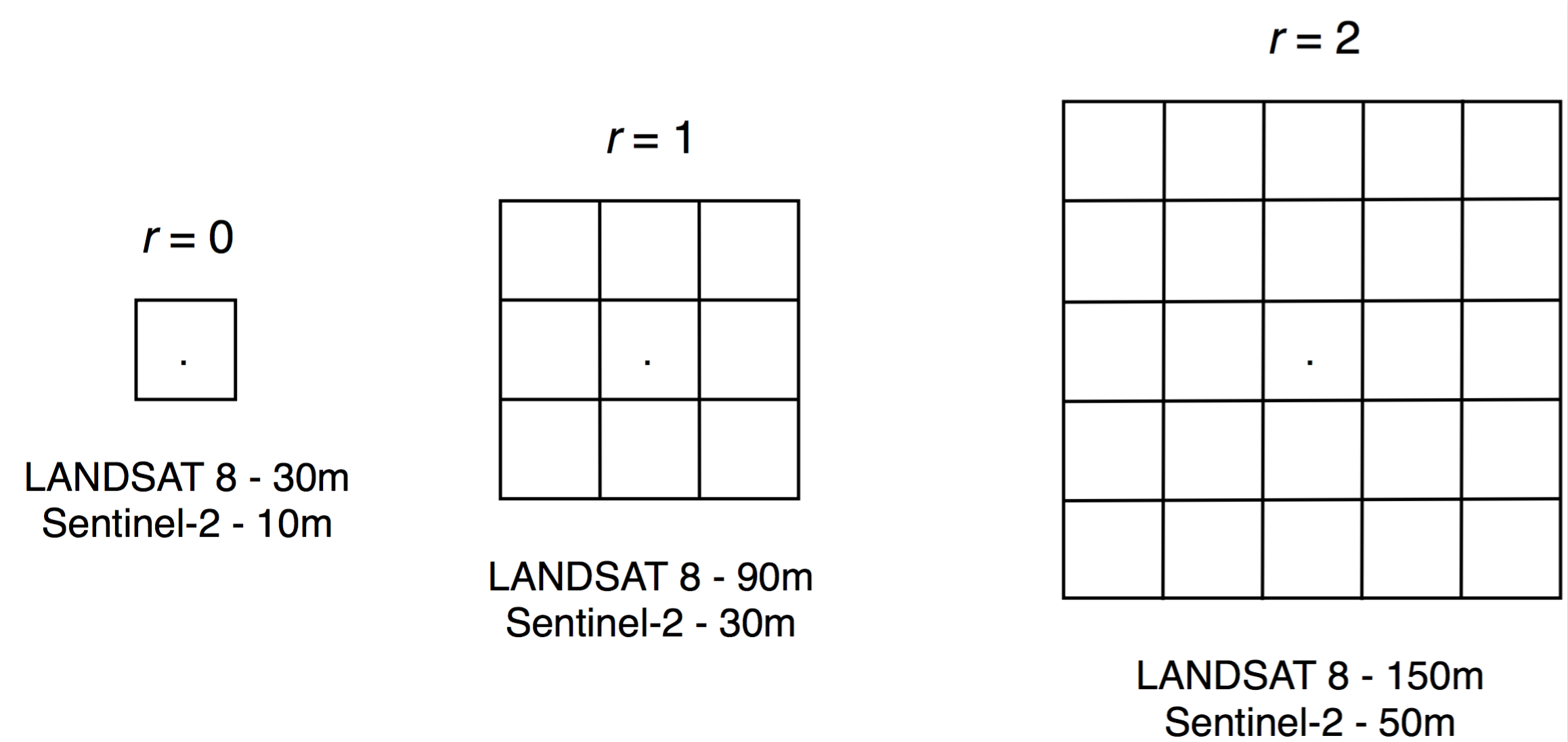}
\caption{The spatial grid around the modelled pixel for different $r$.}
\end{figure}

Thus, we may represent the non-static parameters within each region as  
\begin{linenomath}
\begin{align}
\begin{split}
\textbf{H} &= \left[H_0, H_1, \cdots, H_{N_r-1} \right] \\
\textbf{B}_0 &= \left[B_{0,0}, B_{0,1}, \cdots, B_{0,N_r-1} \right] \\
\textbf{B}_1 &= \left[B_{1,0}, B_{1,1}, \cdots, B_{1,N_r-1} \right] \\
& \vdots \\
\textbf{B}_{N_b-1} &= \left[B_{N_b-1,0}, B_{N_b-1,1}, \cdots, B_{N_b-1,N_r-1} \right] \\
\textbf{q}_0 &= \left[q_{0,0}, q_{0,1}, \cdots, q_{0,N_r-1} \right] \\
\textbf{q}_1 &= \left[q_{1,0}, q_{1,1}, \cdots, q_{1,N_r-1} \right] \\
& \vdots \\
\textbf{q}_{N_b-1} &= \left[q_{N_b-1,0}, q_{N_b-1,1}, \cdots, q_{N_b-1,N_r-1} \right] \\
\end{split}\label{spatial_ext}
\end{align}
\end{linenomath}

Additionally, in many instances it is reasonable to constrain the variance of \textbf{H} 
within the spatial region. That is, a constraint of the form 
$\sigma\left(\textbf{H}\right) < \kappa\, \overbar{\textbf{H}},$ with $\kappa = 0.1$. Where 
$\sigma\left(\textbf{H}\right)$ and $\overbar{\textbf{H}}$ is the standard deviation and mean 
of \textbf{H} respectively. (Alternatively, $\kappa$ can be dependent on depth.) The 
error metric we use in \textit{photic} for the continuity of \textbf{H} is given by 
\begin{linenomath}
\begin{equation}
E_{\textbf{H}} = \sqrt{\frac{1}{N_r}\sum\limits_{i=0}^{N_r-1} 
\begin{cases}
\left(\frac{H_i - \overbar{\textbf{H}}}{\overbar{\textbf{H}}}\right)^2 & \left| H_i - \overbar{\textbf{H}} \right| > \kappa\, \overbar{\textbf{H}}\\
0 & \text{otherwise}
\end{cases}
}.\label{error_H}
\end{equation}
\end{linenomath}

	\subsection{Temporal extension}

We can further increase the spectral data available to the model by using a time series of 
satellite scenes. For LANDSAT-8 and Sentinel-2 scenes, a large number of scenes are 
available and most are usable for our purposes. \\

We assume continuity of the seabed over time. Thus, across multiple scenes the depth, $H$, 
is static up to tidal effects. For $N_s$ scenes, it is assumed we know the tidal corrections 
$H_{\text{tide}_0}$, $H_{\text{tide}_1}$, $\cdots$, $H_{\text{tide}_{N_s-1}}$ to a 
common datum, such that 
\begin{linenomath}
\begin{equation*}
H + H_{\text{tide}_0} = H + H_{\text{tide}_1} = \cdots = H + H_{\text{tide}_{N_s-1}}.
\end{equation*}
\end{linenomath}

We further assume the bottom reflectance is constant over time. This may be a more tenuous 
assumption than our other assumptions, for example for some coral reefs or seagrass beds. \\

Thus, the temporal constraints are $H$, $B_0$, $B_1$, $\cdots$, $B_{N_b-1}$, $q_0$, $q_1$, 
$\cdots$ and $q_{N_b-1}$ are constant, while allowing the water turbidity to vary over 
time - $P$, $G$, $X$ (and $\Delta$), which we represent as 
\begin{linenomath}
\begin{align}
\begin{split}
\textbf{P} &= \left[P_0, P_1, \cdots, P_{N_s-1} \right] \\
\textbf{G} &= \left[G_0, G_1, \cdots, G_{N_s-1} \right] \\
\textbf{X} &= \left[X_0, X_1, \cdots, X_{N_s-1} \right] \\
\mathbf{\Delta} &= \left[\Delta_0, \Delta_1, \cdots, \Delta_{N_s-1} \right].
\end{split}\label{temporal_ext}
\end{align}
\end{linenomath}

These constraints complement the spatial constraints, in that previously $H$, $B_0$, $B_1$, 
$\cdots$, $B_{N_b-1}$, $q_0$, $q_1$, $\cdots$, $q_{N_b-1}$ could vary spatially, while 
$P$, $G$, $X$ are spatially static. 

	\subsection{The multi-spatial, multi-temporal, error metric}

We will now combine the spatial and temporal extensions into a single error metric for our 
model. As before, let $N_s$ be the number of satellite scenes and $N_r$ the size of the spatial 
region around the spectra we are modelling. Then denote $R_{rs_{{\text{measured}}_i}}^j(\lambda)$ 
and $R_{rs_{{\text{modelled}}_i}}^j(\lambda)$ to be the measured and modelled RRS at spatial 
location $i$ (for $i=0,\cdots,N_r-1$) and scene $j$ (for $j=0,\cdots,N_s-1$) for wavelength,
$\lambda$, respectively. Then the RMS error across all scenes and the entire region is 
given by
\begin{linenomath}
\begin{multline}
E_{R_{rs}}^{\text{RMS}}\left(\textbf{P}, \textbf{G}, \textbf{X}, \textbf{B}_0, \textbf{B}_1, \cdots, \textbf{B}_{N_b-1}, \textbf{q}_0, \textbf{q}_1, \cdots, \textbf{q}_{N_b-1}, \textbf{H}, \mathbf{\Delta}\right) = \\
\frac{\sqrt{\sum\limits_{i=0}^{N_r-1}\sum\limits_{j=0}^{N_s-1}\sum\limits_{\lambda} \left(R_{rs_{{\text{modelled}}_i}}^j(\lambda) - R_{rs_{{\text{measured}}_i}}^j(\lambda)\right)^2}}
	{\sum\limits_{i=0}^{N_r-1}\sum\limits_{j=0}^{N_s-1}\sum\limits_{\lambda} R_{rs_{{\text{measured}}_i}}^j(\lambda)},\label{error_RRS}
\end{multline}
\end{linenomath}

\medskip

\noindent similarly, the SAM error is given by

\medskip

\begin{linenomath}
\begin{multline}
E_{R_{rs}}^{\text{SAM}}\left(\textbf{P}, \textbf{G}, \textbf{X}, \textbf{B}_0, \textbf{B}_1, \cdots, \textbf{B}_{N_b-1}, \textbf{q}_0, \textbf{q}_1, \cdots, \textbf{q}_{N_b-1}, \textbf{H}, \mathbf{\Delta}\right) =  \\
\frac{1}{N_r N_s}\sum\limits_{i=0}^{N_r-1}\sum\limits_{j=0}^{N_s-1} \cos^{-1}\left( \frac{R_{rs_{{\text{modelled}}_i}}^j \cdot R_{rs_{{\text{measured}}_i}}^j}
	{\left(R_{rs_{{\text{modelled}}_i}}^j \cdot R_{rs_{{\text{modelled}}_i}}^j\right) \left(R_{rs_{{\text{measured}}_i}}^j \cdot R_{rs_{{\text{measured}}_i}}^j\right)} \right),\label{error_SAM}
\end{multline}
\end{linenomath}

\medskip

\noindent where \textbf{H}, $\textbf{B}_0$, $\textbf{B}_1$, $\cdots$, $\textbf{B}_{N_b-1}$, $\textbf{q}_0$, 
$\textbf{q}_1$, $\cdots$, $\textbf{q}_{N_b-1}$ are given in (\ref{spatial_ext}), and \textbf{P}, 
\textbf{G}, \textbf{X}, $\mathbf{\Delta}$ are given in (\ref{temporal_ext}). We denote $R_{rs_{{\text{modelled}}_i}}^j(\lambda)$ 
as shorthand for 
$R_{rs_{{\text{modelled}}_i}}^j(\textbf{P},$ $\textbf{G},$ $\textbf{X},$ $\textbf{B}_0,$ $\textbf{B}_1,$ $\cdots,$ $\textbf{B}_{N_b-1},$ $\textbf{q}_0,$ $\textbf{q}_1,$ $\cdots,$ $\textbf{q}_{N_b-1},$ $\textbf{H},$ $\mathbf{\Delta},$ $\lambda)$. \\ 

We have three error metrics - the RMS error (\ref{error_RRS}), the SAM error 
(\ref{error_SAM}), and the continuity of \textbf{H} error (\ref{error_H}). We combine 
these into a final, weighted, error metric for our model, 
\begin{linenomath}
\begin{equation}
E_{\text{photic}} = \omega_0 \, E_{R_{rs}}^{\text{RMS}} E_{R_{rs}}^{\text{SAM}} + \omega_1 \, E_{H},
\end{equation}
\end{linenomath}
where $\omega_0 \gg \omega_1$ and $\omega_0 + \omega_1 = 1$. In \textit{photic}, we experimentally 
found $\omega_0 = 0.85$ and $\omega_1 = 0.15$ balances both terms in the error metric. \\

In this formulation of the model, we have $N_r$ parameters for \textbf{H}; $N_b\,N_r$ parameters 
for $\textbf{B}_0$, $\textbf{B}_1$, $\cdots$, $\textbf{B}_{N_b-1}$; $N_b\,N_r$ parameters for 
$\textbf{q}_0$, $\textbf{q}_1$, $\cdots$, $\textbf{q}_{N_b-1}$; and $4 N_s$ parameters for 
\textbf{P}, \textbf{G}, \textbf{X}, $\mathbf{\Delta}$. In total, we have 
\begin{linenomath}
\begin{equation*}
N_{\text{total}} = N_r + 2\,N_b\,N_r + 4 N_s
\end{equation*}
\end{linenomath}
parameters to determine. \\

These spatial and temporal extensions to the model of Lee et al greatly increase data used 
in the spectral matching. Below we compare the number of measured spectra (assuming 
we use the LANDSAT 8 satellite with the coastal, blue, green and red spectral bands) 
across each scene and region to the number of unknown parameters in our model, where 
the number of bottom types, $N_b$ is 2.

\begin{table}[H]
\centering
\caption{In the table on the left we list the number of (simultaneous) \textit{equations} 
for the LANDSAT 8 satellite; and in the table on the right we list the number of model 
unknowns. Both of these are a function of the number of satellite scenes ($N_s$), the 
region size ($N_r$), and the number of bottom types ($N_b$), which for this table 
$N_b = 2$.}
\centerline{
$
  \begin{array}{cc|ccc}
    &\multicolumn{1}{c}{} & \multicolumn{3}{c}{N_r} \\
    &N_{\text{landsat 8}}& 1 & 9 & 25 \\
    \cline{2-5}
    & 1 & 4 & 36 & 100 \\
    \smash{\rotatebox[origin=c]{90}{$N_s$}} & 2 & 8 & \textbf{72} & 200 \\
    & 3 & 12 & 108 & 300 \\
    & 4 & 16 & 144 & 400 
  \end{array}
$ \qquad $
  \begin{array}{cc|ccc}
    &\multicolumn{1}{c}{} & \multicolumn{3}{c}{N_r} \\
    &N_{\text{total}}& 1 & 9 & 25 \\
    \cline{2-5}
    & 1 & 9 & 49 & 129 \\
    \smash{\rotatebox[origin=c]{90}{$N_s$}} & 2 & 13 & \textbf{53} & 133 \\
    & 3 & 17 & 57 & 137 \\
    & 4 & 21 & 61 & 141  
  \end{array}
$}
\end{table}

\medskip

\noindent A common setting for our modelling is $N_s = 2$, $N_r = 9$ and $N_b = 2$, which is highlighted above. 

\section{Inverting the Model}

We will now detail how we invert (\ref{rrs_model}) to obtain estimates for \textbf{P}, 
\textbf{G}, \textbf{X}, $\textbf{B}_0$, $\textbf{B}_1$, $\cdots$, $\textbf{B}_{N_b-1}$, 
\textbf{H}, $\textbf{q}_0$, $\textbf{q}_1$, $\cdots$, $\textbf{q}_{N_b-1}$ and 
$\mathbf{\Delta}$ by finding a (global) minimum of $E_{\text{photic}}$. 
 
	\subsection{Initial depth estimate}\label{lyzenga_method}

Lee et al \cite{lee1999} used $H = 10.0$ m as an initial depth estimate and subsequently 
used an estimate of $H = 1/(6P)$ \cite{dekker2011}. Starting the model inversion at a 
reasonable depth estimate was a technique used by Albert et al \cite{albert2006}. \\

If reliable depth profiles are known, then initial, and often accurate, depth estimates 
can be obtained using an empirical method \cite{polycn1976}\cite{lyzenga1978}\cite{lyzenga2006}. 
Closely following the model of Lyzenga et al \cite{lyzenga2006} - the empirical depth, 
$H_{\text{empirical}}$, for a $n$-band spectrum is given by 
\begin{linenomath}
\begin{equation*}
H_{\text{empirical}}\left(h_1,h_2,\cdots,h_n\right) = h_0 - \sum_{i=1}^{n} h_i \log(r_{rs}(\lambda_{i-1}) - r_{rs_{\infty}}(\lambda_{i-1})),
\end{equation*}
\end{linenomath}
where $h_0 = \sum\limits_{i=1}^n h_i \log(\rho(\lambda_{i-1})/\pi)$ and $\sum\limits_{i=1}^n h_i\, k(\lambda_{i-1}) = 1$.\\

The subsurface RRS of optically deep waters is estimated using the method of Lyzenga 
et al \cite{lyzenga2006}. We compute both the mean subsurface RRS of optically deep 
water, $\overbar{r}_{rs_\infty}(\lambda)$, and the standard deviation of the subsurface 
RRS of optically deep water, $\sigma\left(r_{rs_\infty}(\lambda)\right)$ in each 
band. It is worth noting that this is a scene-wide estimate. \\

The attenuation coefficients are estimated using the blue and green spectral bands  
\begin{linenomath}
\begin{equation*}
\frac{k(\text{blue})}{k(\text{green})} \approx \frac{\log\left(r_{rs}(\text{blue}) - r_{rs_{\infty}}(\text{blue})\right)}{\log\left(r_{rs}(\text{green}) - r_{rs_{\infty}}(\text{green})\right)},
\end{equation*}
\end{linenomath}
then interpolate across spectra and water types from a table of spectral attenuation 
coefficients for different coastal and oceanic water types to directly estimate the 
attenuation coefficients and the water type \cite{jerlov1968}\cite{morel1998}. Again, 
it is worth noting that this is a scene-wide estimate. \\

Consider $N$ soundings $s_0, s_1, \cdots, s_{N-1}$, we begin by generating weights 
for each soundings $w_0, w_1, \cdots, w_{N-1}$ such that 
\begin{linenomath}
\begin{equation*}
w_i = 1 - W_i/M, \quad \text{where} \quad W_i = \sum\limits_{j=0}^{N-1} e^{-\left(s_i - s_j\right)^2},
\end{equation*}
\end{linenomath}

\noindent where $M = \max(W_0, W_1, \cdots, W_{N-1})$. This weighting scheme 
prevents frequently occurring depths from skewing the subsequence fitting. 
We construct a weighted, root-mean-square, relative error metric, 
$E_{\text{empirical}}$, such that 
\begin{linenomath}
\begin{equation*}
E_{\text{empirical}}\left(h_1, h_2, \cdots, h_n\right) = 
\sqrt{\frac{\sum\limits_{i=0}^{N-1} w_i \left(\frac{H_{\text{empirical}_i} - s_i}{s_i}\right)^2}{\sum\limits_{i=0}^{N-1} w_i}},
\end{equation*}
\end{linenomath} 
where $H_{\text{empirical}_i}$ is shorthand for 
$H_{\text{empirical}_i}\left(h_1,h_2,\cdots,h_n\right)$ at the spatial location 
of the sounding, $s_i$. \\

To find the minima of $E_{\text{empirical}}$, we use multiple iterations of the 
simplex algorithm \cite{oneil1971}, each with a pseudo-randomly generated initial 
simplex. This random initialisation with multiple iterations helps prevent the 
optimisation from settling in a local minimum.  \\

With multiple scenes, the depths from each scene from the empirical algorithm can 
be synthesised into a single depth estimate by way of a temporal median across 
all modelled scenes (once we account for tidal variations). Alternatively, a 
weighted mean can be used. In this case, the weights are inversely proportional 
to the final $E_{\text{empirical}}$ of each scene. 

	\subsection{Ordering pixels by spectral angle}

It would be natural within a computer implementation to model the pixels in a column- or 
row-major order. However in \textit{photic}, pixels within the modelling domain are 
sorted by spectral angle above the deep water mean (via the SAM). The spatial indices 
of the pixels are stored so we can return the modelling results to their original 
location. The model starts with the pixels closest to the mean deep water spectrum. 
We will subsequently see why this ordering is favourable to our modelling. 

	\subsection{Initial estimates for \textbf{P}, \textbf{G}, \textbf{X} and $\mathbf{\Delta}$}\label{estimates_1}

The initial estimates closely follows Lee et al \cite{lee1999}. Without any knowledge of field 
samples, the initial parameterisation of \textbf{P}, \textbf{G}, \textbf{X} and $\mathbf{\Delta}$ 
is given by 
\begin{linenomath}
\begin{align*}
P_j &= 0.072 \left({\overbar{R}_{rs}}^j(440)/{\overbar{R}_{rs}}^j(550)\right)^{-1.7}\\
G_j &= 1.5 P_j \\
X_j &= 30\, a_w(640) {\overbar{R}_{rs}}^j(640) \\
\Delta_j &= {\overbar{R}_{rs}}^j(750)
\end{align*}
\end{linenomath}
where $\overbar{R}_{rs}$ is the average over the $N_r$ spectra in the region. We call this 
parameterisation a \textit{cold start}, as previously modelled pixels with similar spectra have 
not guided the current starting point. \\

Alternatively, as the pixels are sorted by their SAM from the deep water mean RRS, if the 
current region of pixels being modelled is within a model-defined threshold of the 
previously modelled pixel, then we \textit{hot start} the model with the previous optimal 
parameterisation for \textbf{P}, \textbf{G}, \textbf{X} and $\mathbf{\Delta}$. This 
significantly reduces the number of iterations before convergence. 

	\subsection{Initial estimates for \textbf{B} and \textbf{q}}\label{estimates_2}

The initial estimate of $\textbf{B} = 0.4 {R_{rs_i}}(490)$ follows the HOPE model of Lee 
et al \cite{dekker2011}. We arbitrarily set $q_i = 1.0$, which corresponds to equal 
proportions of each bottom type. Unlike the initialisation of \textbf{P}, \textbf{G}, 
\textbf{X} and $\mathbf{\Delta}$, we do not hot start \textbf{B} and $\textbf{q}$; 
as this would make assumptions on the continuity of the bottom reflectance. 

	\subsection{The optimisation approach}

With our initial estimation of all parameters (both spatially and temporally) in (\ref{rrs_model}), we 
now seek the parameterisation which minimises our error metric, $E_{\text{photic}}$. \textit{photic} 
uses the simplex algorithm \cite{oneil1971} to perform the optimisation. This algorithm is 
derivative free and converges quickly if a reasonable starting point (simplex) is chosen. \\

When the model is cold started and an initial estimate for \textbf{H} is not given, then we perform 
multiple optimisations with starting points at the following depths 0.1, 0.5, 1.0, 1.5, 2.0, 3.0, 
4.0, 5.0, 6.0, 8.0, 10.0, 12.5, 15.0, 17.5, 20.0, 25.0, 30.0 m. This is computationally expensive, 
but rare as most pixels are hot-started. \\

\textit{photic} can also take a range for each parameter. Within the simplex optimisation each 
parameter is constrained to be within its user-defined range. These ranges could come from 
local knowledge or from physical measurements from the modelling domain. 

	\subsection{The dynamic lookup table approach}

Modelling large areas using the optimisation approach requires significant computing resources. 
One way to decrease this computational burden is the inclusion of a small, dynamic, lookup table 
(LUT) which is searched prior to running the optimisation approach. If the match between the 
current pixel and spectra in the LUT (in the sense of the SAM between the two spectra) is 
below a model-defined threshold then the LUT is used and the optimisation approach is skipped. \\

Whenever a pixel is modelled with the optimisation approach, if $E_{\text{photic}}$ is below a 
model-defined threshold then the result of this modelling is stored in the LUT along with a 
timestamp. When a new entry is stored in the LUT, the oldest entry is discarded (overwritten). 
In \textit{photic} the LUT is deliberately small, with only 256 entries. This way searching 
the LUT is fast, however the LUT should still contain many spectra relevant to the current 
pixel being modelled as the pixels are sorted prior to modelling. \\

In \textit{photic} the threshold on entering a modelled pixel to the LUT is initially 
$E_{\text{photic}} < \max\left(1.5, 1.125\,N_s\right)$, but can be adaptively modified. The 
threshold cannot exceed $2.5 + 2.5 N_s$ (where the units of $E_{\text{photic}}$ is 
relative percentage error). This prevents poorly modelled pixels from further degrading 
pixels within the modelling domain. \\

Using a dynamic LUT in conjunction with sorting the spectra results in significant speed 
improvements. Generally, more than 95\% of pixels are modelled using the LUT. The LUT can 
model in excess of $100\,000\,\text{px}/\text{sec}$, which is at least three orders of 
magnitude faster than the optimisation approach. 

	\subsection{An example inversion}

As a detailed example of the model, we give a breakdown of the optimisation 
process for $N_s = 2$ and $N_r = 9$. The spectra are taken from the case study in 
section \ref{case_study}, for scenes \texttt{LC81150752018058LGN00} and 
\texttt{LC81150752019253LGN00}. The location of these spectra are 114.361135E, 
21.676824S and from the sounding data the water depth is 14.9 m LAT. The bottom 
of atmosphere RRS spectra for each scene and region is given in the table below. 

\begin{table}[H]
\centering
\caption{RRS spectra ($\times 10^3$) for two satellite scenes ($N_s=2$) and 9 
neighbourhood samples ($N_r = 9$, that is one central spectra and 8 surrounding 
spectra).}
$
\begin{array}{l c c c c | c c c c}
& & \text{scene 1} & &  & & \text{scene 2} & &  \\
\lambda(\text{nm}) & 443     &     483 &     561 &     655 &     443 &     483 &     561 &     655 \\
\cline{2-9}
& 7.9768 & 9.6534 & 6.8729 & 1.7796 & 8.5987 & 9.9610 & 6.3355 & 1.6588 \\ 
& 8.0830 & 9.8336 & 6.9159 & 1.9415 & 8.7791 & 9.9258 & 6.3835 & 1.6932 \\
& 8.3104 & 9.9590 & 7.2173 & 2.1246 & 8.7831 & 9.9830 & 6.4267 & 1.6760 \\
& 7.9996 & 9.7127 & 6.8662 & 1.7158 & 8.5305 & 10.0314 & 6.3499 & 1.7493\\
\smash{\rotatebox[origin=c]{90}{regions}} & 8.2061 & 9.8861 & 6.9027 & 1.8989 & 8.7230 & 9.9698 & 6.3835 & 1.7407 \\
& 8.4108 & 10.0069 & 7.2339 & 2.0778 & 8.8393 & 10.0094 & 6.4747 & 1.6976 \\
& 8.2990 & 9.9567 & 7.1643 & 1.9415 & 8.7070 & 10.0666 & 6.5420 & 1.8614 \\
& 8.4620 & 10.1141 & 7.1974 & 2.1246 & 8.6949 & 10.0314 & 6.5612 & 1.8010 \\
& 8.4260 & 10.0639 & 7.2140 & 2.0906 & 8.7992 & 10.0182 & 6.5660 & 1.7407 \\
\end{array}
$
\end{table}

All parameters were initialised using the scheme described in sections \ref{lyzenga_method}, 
\ref{estimates_1} and \ref{estimates_2}.

\begin{figure}[H]
\centering
\includegraphics[width=0.8\textwidth]{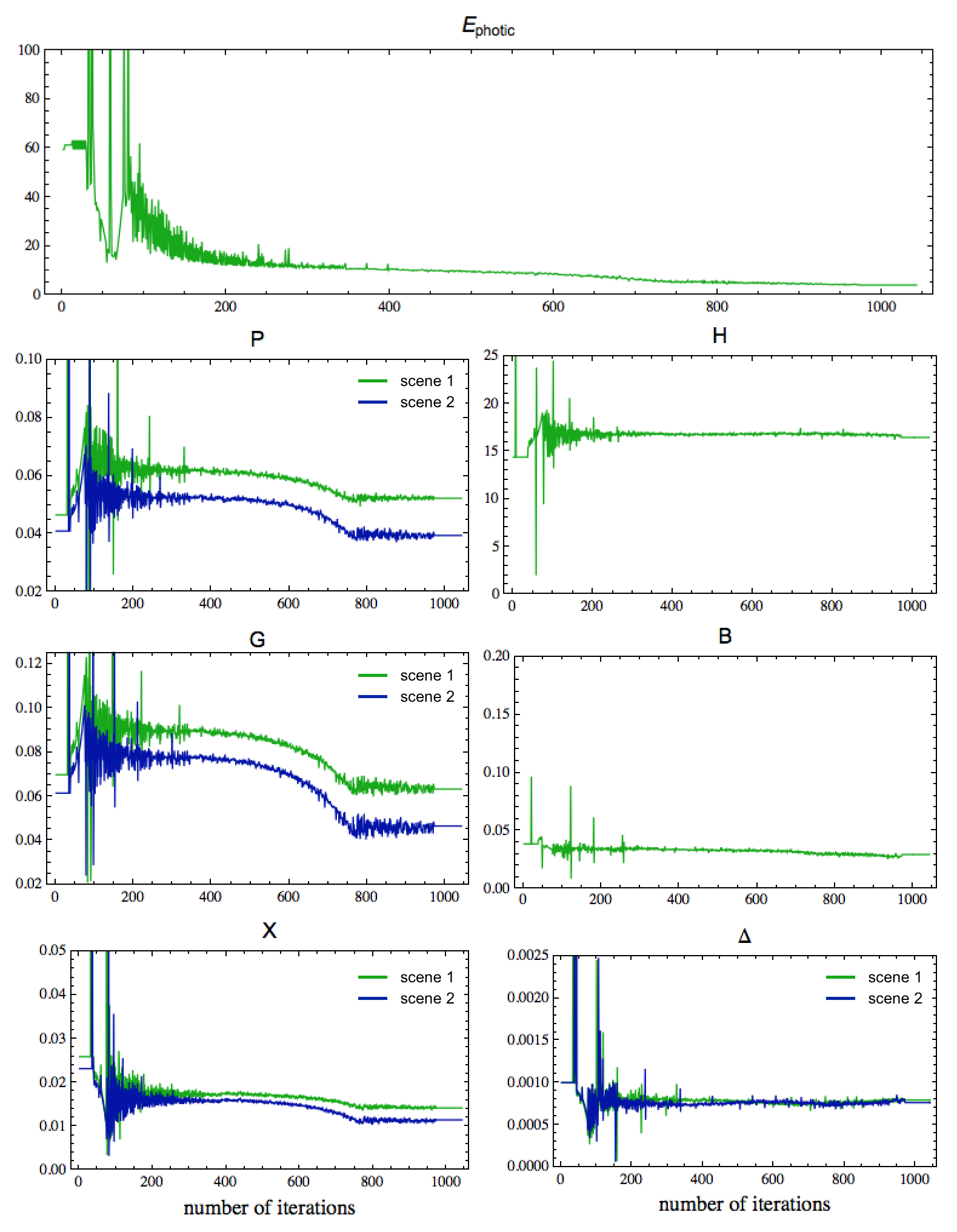}
\caption{A plot of the path taken by each model variable during the minimisation process at the 
central modelled spectra.}
\end{figure}

\begin{figure}[H]
\centering
\includegraphics[width=1.0\textwidth]{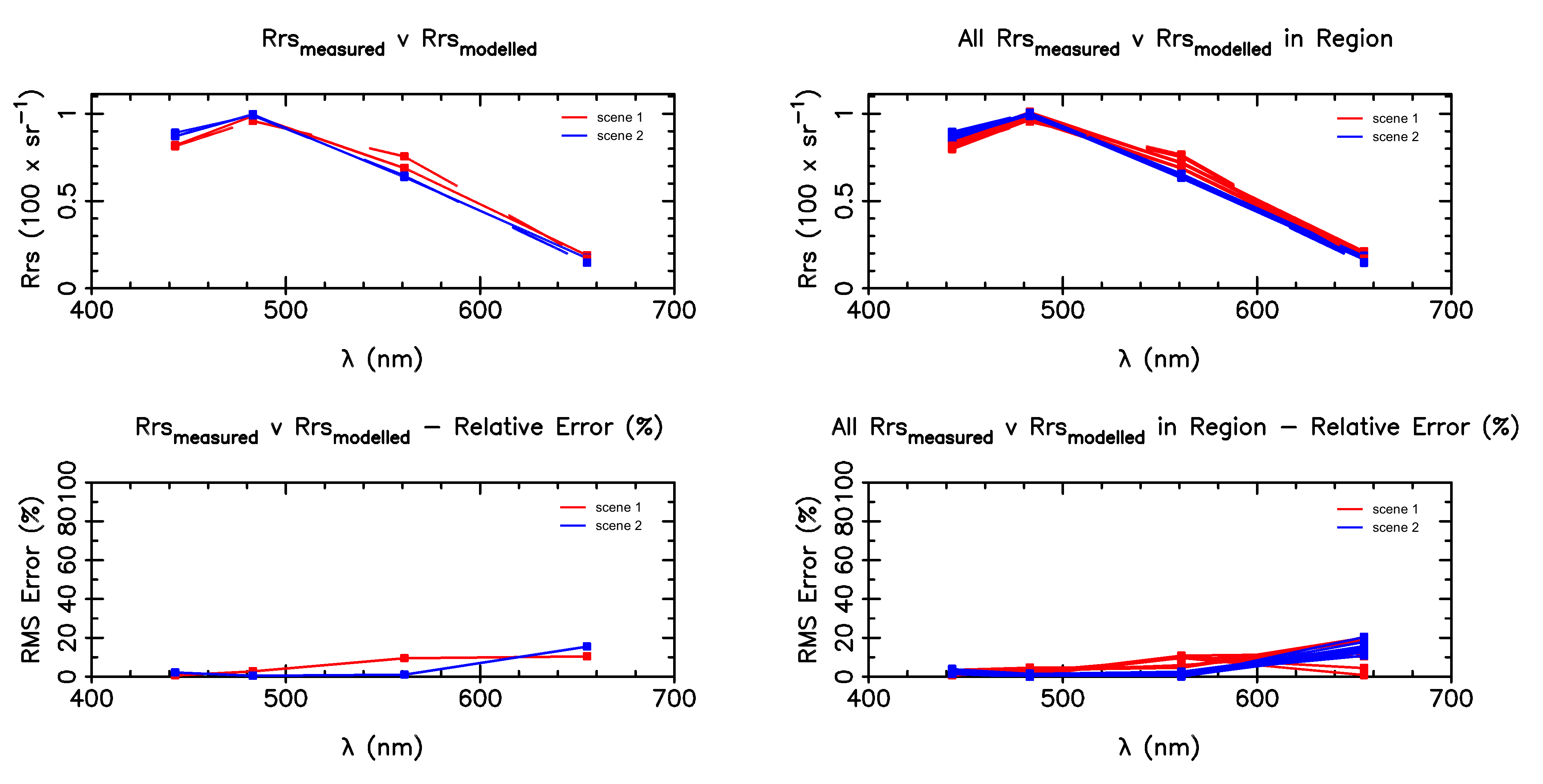}
\caption{A plot comparing $R_{rs_{\text{measured}}}$ with $R_{rs_{\text{modelled}}}$ at 
the central modelled spectra and at all $N_r$ surrounding spectra in the region.}
\end{figure}

In summary, the model required 1043 iterations to converge to a model error of 
$E_{\text{photic}} = 3.96\%$. The derived parameters were 
\begin{linenomath}
\begin{align*}
\textbf{P} & = \left[0.05226967, 0.03944374 \right] \\
\textbf{G} & = \left[0.06326824, 0.04656528 \right] \\
\textbf{X} & = \left[0.01420787, 0.01148168 \right] \\
\mathbf{\Delta} & = \left[0.00079306, 0.00076364 \right] \\
\textbf{B} & = \left[0.0248, 0.0295, 0.0287, 0.0299, 0.0296, 0.0293, 0.0264, 0.0264, 0.0290 \right] \\ 
\textbf{H} & = \left[16.12, 18.52, 17.37, 15.20, 16.47, 18.37, 18.49, 17.74, 18.32 \right]
\end{align*}
\end{linenomath}
Thus, the model-derived depth is 16.47 m. 

\section{Iterative Estimation of Depth, Water Turbidity and Bottom Composition}

	\subsection{The first approximation}\label{first_approximation}

In the first approximation to the inversion, we model all individual scenes, pairs of 
scenes, 3-tuples of scenes and 4-tuples of scenes. In each of these modelling iterations, 
all parameters \textbf{P}, \textbf{G}, \textbf{X}, $\textbf{B}_0$, $\textbf{B}_1$, 
$\cdots$, $\textbf{B}_{N_b-1}$, $\textbf{q}_0$, $\textbf{q}_1$, $\cdots$, 
$\textbf{q}_{N_b-1}$, \textbf{H} and $\mathbf{\Delta}$ are estimated by the model and 
stored for subsequent use. 

	\subsection{Depth averaging}

If reliable depth profiles are not known, then we take the median of all 
iterations of the singletons, pairs, 3- and 4-tuples from the first approximation 
above. We use a weighted median, where the weight is proportional to the number of 
scenes in the model. That is, proportional to $N_s$. 

	\subsection{Aligning $H$, with known depth profiles}

If reliable depth profiles are known, then the model-derived depths, $H$, can be 
\textit{aligned} with these profiles. This technique, for a single scene, was used 
by Ohlendorf et al \cite[Figure 9. Depth validation at Rottnest 
Island, 2005]{ohlendorf2011} as a post processing step to the model inversion. \\

From the $n$ model-derived depths, $H_0, H_1, \cdots, H_{n-1}$, we construct a 
single depth $H_{\text{aligned}}$, such that 
\begin{linenomath}
\begin{equation*}
H_{\text{aligned}}(c_0, c_1, \cdots, c_{n-1}, a_0, a_1, \cdots a_{n-1}, b_0, b_1, \cdots, b_{n-1}) = \frac{\sum\limits_{i=0}^{n-1} c_i a_i H_i^{b_i}}{\sum\limits_{i=0}^{n-1}c_i},
\end{equation*}
\end{linenomath}
where $c_i,a_i,b_i > 0$. We construct the same weighting scheme as used in the empirical 
method - consider $N$ soundings $s_0, s_1, \cdots, s_{N-1}$, we begin by generating 
weights for each soundings $w_0, w_1, \cdots, w_{N-1}$ such that 
\begin{linenomath}
\begin{equation*}
w_i = 1 - W_i/M, \quad \text{where} \quad W_i = \sum\limits_{j=0}^{N-1} e^{\left(s_i - s_j\right)^2},
\end{equation*}
\end{linenomath}
where $M = \max(W_0, W_1, \cdots, W_{N-1})$. We construct a weighted, root-mean-square, 
relative error metric, $E_{\text{aligned}}$, such that 
\begin{linenomath}
\begin{equation*}
E_{\text{aligned}}(c_0, c_1, \cdots, c_{n-1}, a_0, a_1, \cdots a_{n-1}, b_0, b_1, \cdots, b_{n-1}) = 
\sqrt{\frac{\sum\limits_{i=0}^{N-1} w_i \left(\frac{H_{\text{aligned}_i} - s_i}{s_i}\right)^2}{\sum\limits_{i=0}^{N-1} w_i}}
\end{equation*}
\end{linenomath}
where $H_{\text{aligned}_i}$ is shorthand for $H_{\text{aligned}}(c_0, c_1, \cdots, c_{n-1}, a_0, a_1, \cdots a_{n-1}, b_0, b_1, \cdots, b_{n-1})$ at the spatial location of the sounding, $s_i$. \\  

To find the minima of $E_{\text{aligned}}$, we use the simplex algorithm \cite{oneil1971} with an 
initial simplex of $c_i,a_i,b_i = 1$ for all $i$. If the model is well calibrated, 
then we expect this starting point to be close to the minima. \\


	\subsection{Averaging $k$}

If the modelled attenuation coefficients, $k(\lambda)$ (eqn. \ref{k_eqn}), is relatively 
constant over time, then we can average $k(\lambda)$ over all model iterations from 
\ref{first_approximation}. Otherwise all individual model-derived parameters can be 
averaged. 

	\subsection{Deriving bottom types}

We perform an additional optimisation iteration to improve the accuracy of the unmixing of 
multiple bottom types. In this modelling iteration we use the previously estimated  
\textbf{P}, \textbf{G}, \textbf{X}, $\textbf{B}_0$, $\textbf{B}_1$, $\cdots$, $\textbf{B}_{N_b-1}$, 
\textbf{H} and $\mathbf{\Delta}$. However, \textbf{P}, \textbf{G}, \textbf{X} and 
$\mathbf{\Delta}$ are constant. We use the values of $\textbf{B}_0$, $\textbf{B}_1$, 
$\cdots$, $\textbf{B}_{N_b-1}$ as an initialisation for subsequent modelling. \\

We begin by expressing \ref{rrs_model} in terms of $\rho(\lambda)$, which we 
term $\rho_{\text{modelled}}(\lambda)$ with parameters $P$, $G$, $X$, $H$, $\Delta$
\begin{linenomath}
\begin{multline*}
\rho_{\text{modelled}}(P, G, X, H, \Delta, \lambda) = \pi \exp\left(\left(\frac{1}{\cos(\theta_{\text{sun}})} + \frac{D_B(\lambda)}{\cos(\theta_{\text{view}})}\right)\,k(\lambda)\,H\right) \bigg[r_{rs}(\lambda) - \\
\left. r_{rs_{\infty}}(\lambda) \left(1 - \exp\left(-\left(\frac{1}{\cos(\theta_{\text{sun}})} + \frac{D_C(\lambda)}{\cos(\theta_{\text{view}})}\right)\,k(\lambda)\,H\right)\right) \right].
\end{multline*}
\end{linenomath}
where
\begin{linenomath}
\begin{equation*}
r_{rs}(\lambda) = \frac{2(R_{rs}(\lambda) - \Delta)}{1 + 3(R_{rs}(\lambda) - \Delta)}
\end{equation*}
\end{linenomath}
and the corresponding unmixed $\rho(\lambda)$, which we term 
$\rho_{\text{unmixed}}(B_0, B_1, \cdots, B_{N_b-1}, q_0, q_1, \cdots, q_{N_b-1},\lambda)$ is given by 
\begin{linenomath}
\begin{equation*}
\rho_{\text{unmixed}}(B_0, B_1, \cdots, B_{N_b-1}, q_0, q_1, \cdots, q_{N_b-1}, \lambda) = \frac{\sum\limits_{i=0}^{N_b-1}B_i\,q_i\,\rho_{\text{bottom}_i}(\lambda)}{\sum\limits_{i=0}^{N_b-1}q_i}.
\end{equation*}
\end{linenomath}
Then we use the averaged (or aligned) depths as \textbf{H} across all subsequent model iterations. The 
values of \textbf{P}, \textbf{G}, \textbf{X}, $\mathbf{\Delta}$ are averaged across all previous 
modelling iterations from \ref{first_approximation} for each scene. If noise is present in these datasets 
then we apply a median filter, prior to computing $\rho_{\text{modelled}}$. \\

In this modelling iteration we keep \textbf{P}, \textbf{G}, \textbf{X}, \textbf{H}, $\mathbf{\Delta}$ 
static and optimise only for $\textbf{B}_0$, $\textbf{B}_1$, $\cdots$, $\textbf{B}_{N_b-1}$, 
$\textbf{q}_0$, $\textbf{q}_1$, $\cdots$, $\textbf{q}_{N_b-1}$. As previously, we assume the bottom 
spectra do not change over time, this allows us to simultaneously model multiple scenes. We index 
the $N_s$ scenes, with $j$. \\ 

The error metric for the unmixing is given by 
\begin{linenomath}
\begin{equation*}
E_{\text{unmixed}}\left(\textbf{B}_0, \textbf{B}_1, \cdots, \textbf{B}_{N_b-1}, 
\textbf{q}_0, \textbf{q}_1, \cdots, \textbf{q}_{N_b-1}\right) = 
E^{\text{RMS}}_{\text{unmixed}}\,E^{\text{SAM}}_{\text{unmixed}},
\end{equation*}
\end{linenomath}
where
\begin{linenomath} 
\begin{equation*}
E^{\text{RMS}}_{\text{unmixed}}\left( \textbf{B}, \textbf{q}_0, \textbf{q}_1, \cdots, \textbf{q}_{N_b-1} \right) =
\frac{\sqrt{\sum\limits_{j=0}^{N_s-1}\sum\limits_{\lambda} \left(\rho_{\text{unmixed}}^j - \rho_{\text{modelled}}^j\right)^2}}
	{\sum\limits_{j=0}^{N_s-1}\sum\limits_{\lambda} \rho_{\text{modelled}}^j}
\end{equation*}
\end{linenomath}
and 
\begin{linenomath}
\begin{multline*}
E^{\text{SAM}}_{\text{unmixed}}\left(\textbf{B}_0, \textbf{B}_1, \cdots, \textbf{B}_{N_b-1}, \textbf{q}_0, \textbf{q}_1, \cdots, \textbf{q}_{N_b-1} \right) = \\
\frac{1}{N_s}\sum\limits_{j=0}^{N_s-1} \cos^{-1}\left( \frac{\rho_{\text{unmixed}}^j \cdot \rho_{\text{modelled}}^j}
	{\left(\rho_{\text{unmixed}}^j \cdot \rho_{\text{unmixed}}^j\right) \left(\rho_{\text{modelled}}^j \cdot \rho_{\text{modelled}}^j\right)} \right),
\end{multline*}
\end{linenomath}
where $\rho_{\text{unmixed}}^j$ is shorthand for $\rho_{\text{unmixed}}^j(B_0, B_1, \cdots, B_{N_b-1}, , q_0, q_1, \cdots, q_{N_b-1}, \lambda)$ 
and $\rho_{\text{modelled}}^j$ is shorthand for $\rho_{\text{modelled}}^j(P, G, X, H, \Delta, \lambda)$. 

	\subsection{Depth error estimate}

Obtaining a realistic depth error estimate requires accounting for uncertainties in both 
the model inputs and the model itself. \\

We estimate the sensor noise by computing the RRS standard deviation in optically deep 
water. The model then adds (or subtracts) random noise within the threshold of the 
estimated sensor noise. The corresponding standard deviation of the resulting depths 
over many trials is the depth error estimate. \\

We can further account for errors in the model, bottom reflectance, absorption and 
backscattering spectra by scaling the depth error estimate, if reasonable upper 
bounds on the error estimates of these quantities are known. 

\section{Case Study -- Murion Island Marine Management Area}\label{case_study}

As a small exemplar case study we modelled a 260 $\text{km}^2$ region to the south 
of North Murion Island, including Sunday Island, Combe Reef and Exmouth Reef; which 
are off the Pilbara Coast in Western Australia. This area is part of the Murion 
Island Marine Management Area. It is also an important area for commercial marine 
traffic.

\begin{figure}[H]
\centering
\includegraphics[width=1.0\textwidth]{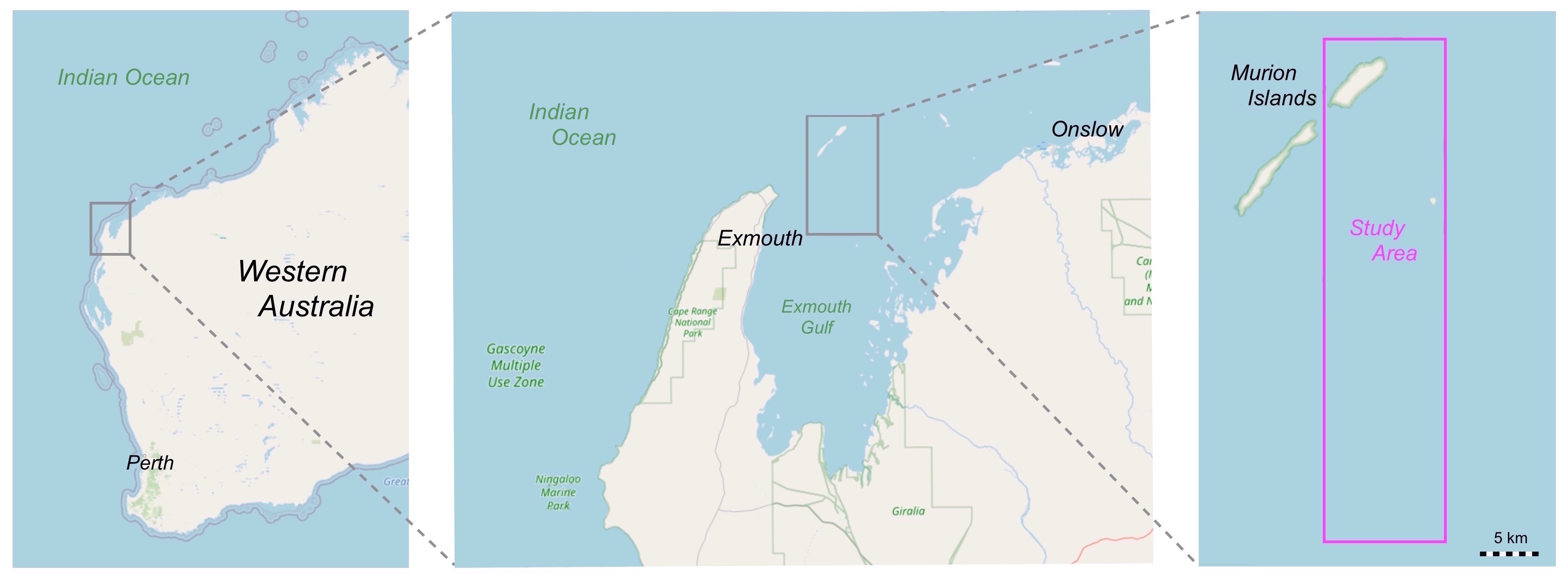}
\caption{A map showing the location of the study area in northern Western Australia \cite{open_street_maps}.}
\end{figure}

A single-beam sonar survey was conducted in 2011 by Transport WA. Within the spatial 
domain of this case study there are 53\,194 sonar measurements, the majority of 
which are between depths of 12 and 22 m. This survey is approximately 32 km by 9.5 km. \\

The satellite imagery used for this study was from the LANDSAT 8 satellite. The scenes from 
this satellite are $170\,\text{km} \times 185\,\text{km}$ in size at a horizontal resolution of 
approximately 30 m. The USGS provides open access to the entire archive of LANDSAT imagery 
dating back to 1972. In total, 4 scenes were selected for modelling

\begin{table}[H]
\centering
\caption{LANDSAT-8 scenes used in the case study.}
\begin{tabular}{ c | c | c | c | c | c }
study ID & 
scene ID & 
\begin{tabular}{@{}c@{}} \text{WRS-2} \\ \text{path} \end{tabular} & 
\begin{tabular}{@{}c@{}} \text{WRS-2} \\ \text{row} \end{tabular} & 
date & 
\begin{tabular}{@{}c@{}} \text{sun} \\ \text{elevation} \end{tabular} \\
\hline
1 & \texttt{LC81150752015210LGN02} & 115 & 75 & 2015-07-29 & 38.86$^{\circ}$ \\
2 & \texttt{LC81150752018058LGN00} & 115 & 75 & 2018-02-27 & 55.22$^{\circ}$ \\ 
3 & \texttt{LC81150752018266LGN00} & 115 & 75 & 2018-09-23 & 54.85$^{\circ}$ \\
4 & \texttt{LC81150752019253LGN00} & 115 & 75 & 2019-09-10 & 50.68$^{\circ}$
\end{tabular}
\end{table}

Atmospheric correction and sun glint correction was performed with ACOLITE \cite{vanhellemont2019}. \\

\begin{figure}[H]
\centering
\includegraphics[width=0.75\textwidth]{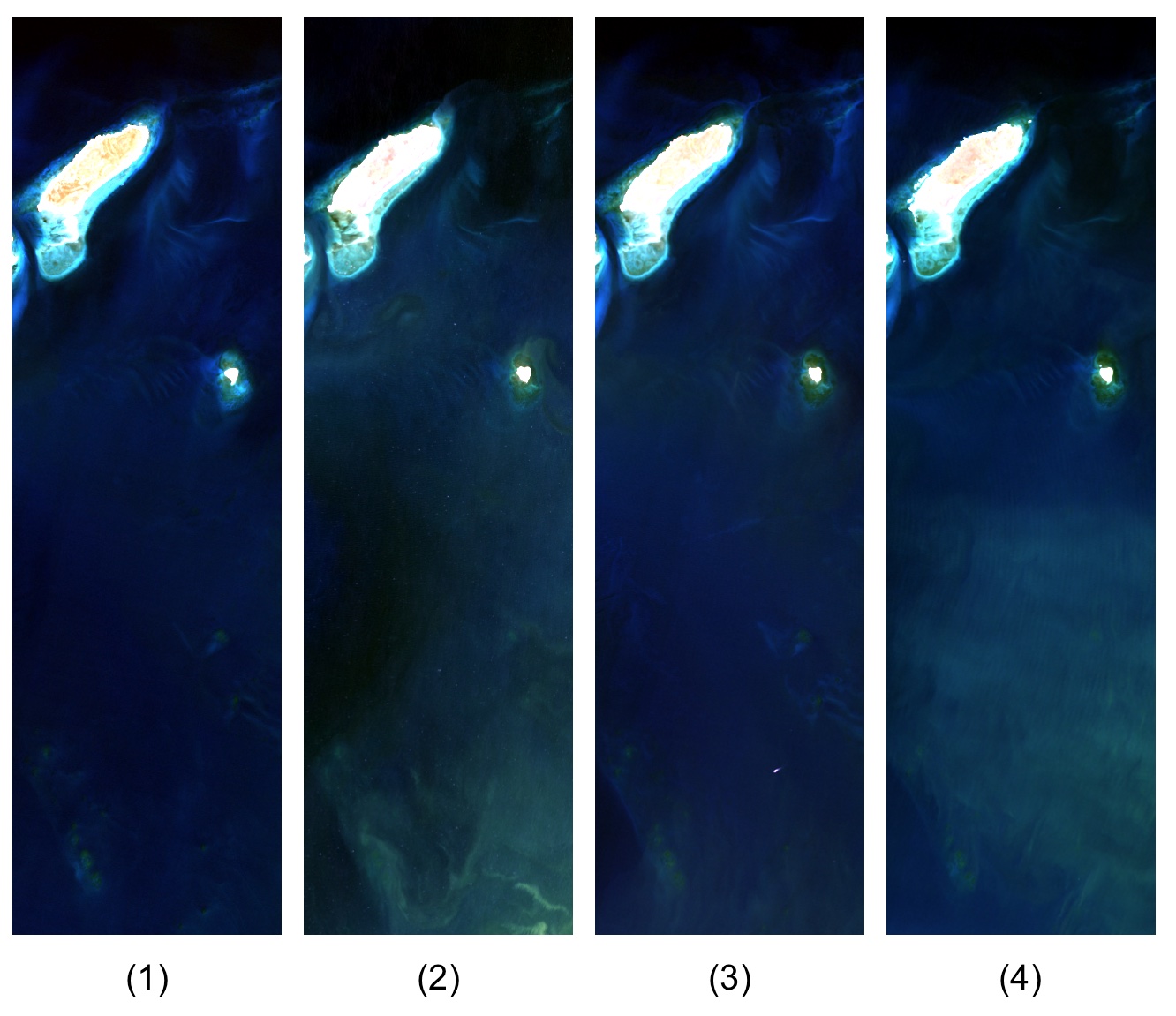}
\caption{Natural colour images of the 4 atmospherically-corrected LANDSAT-8 scenes used in 
this case study; labelled by scene ID. It is visually evident that scenes 2 and 4 exhibit 
higher turbidity in the southern area of the case study. In scene 3 a large vessel is 
present within the AOI.}
\end{figure}

The bottom types used in this study were sand, seagrass and coral. Unfortunately, these 
spectra were not collected from the study site and at best can only be used as 
approximate representative samples. \\ 

Depth soundings used for the initial depth estimation were digitised from chart WA 900 - Point 
Murat to North Murion Island. The empirical model (see section \ref{lyzenga_method}) performed 
well, with a mean relative error of 15\% and a mean absolute error of 1.74 m. \\

\textit{photic} was parameterised with $N_r = 9$, $N_s = 1, 2, 3, 4$, $N_b = 3$. In total, 
we modelled 15 combinations of these 4 scenes. Where the combinations of modelled scenes, 
by study ID, were [1], [2], [3], [4], [1,2], [1,3], [1,4], [2,3], [2,4], [3,4], 
[1,2,3], [1,2,4], [1,3,4], [2,3,4], [1,2,3,4]. The final depth was taken as the spatial 
median of the 15 modelling iterations. \\

All 15 model iterations of \textit{photic} ran sequentially in 5 hours. The lookup table 
accounted for modelling the majority of the points. Without the lookup table the modelling 
would have taken around two weeks (running sequentially on a single core). \\

In the scatter plot below a vertical correction of -0.75 m was applied to the 
model-derived bathymetry. This accounts for tidal variations and a correction to the 
vertical datum of the sonar data. 
\begin{figure}[H]
\centering
\includegraphics[width=0.66\textwidth]{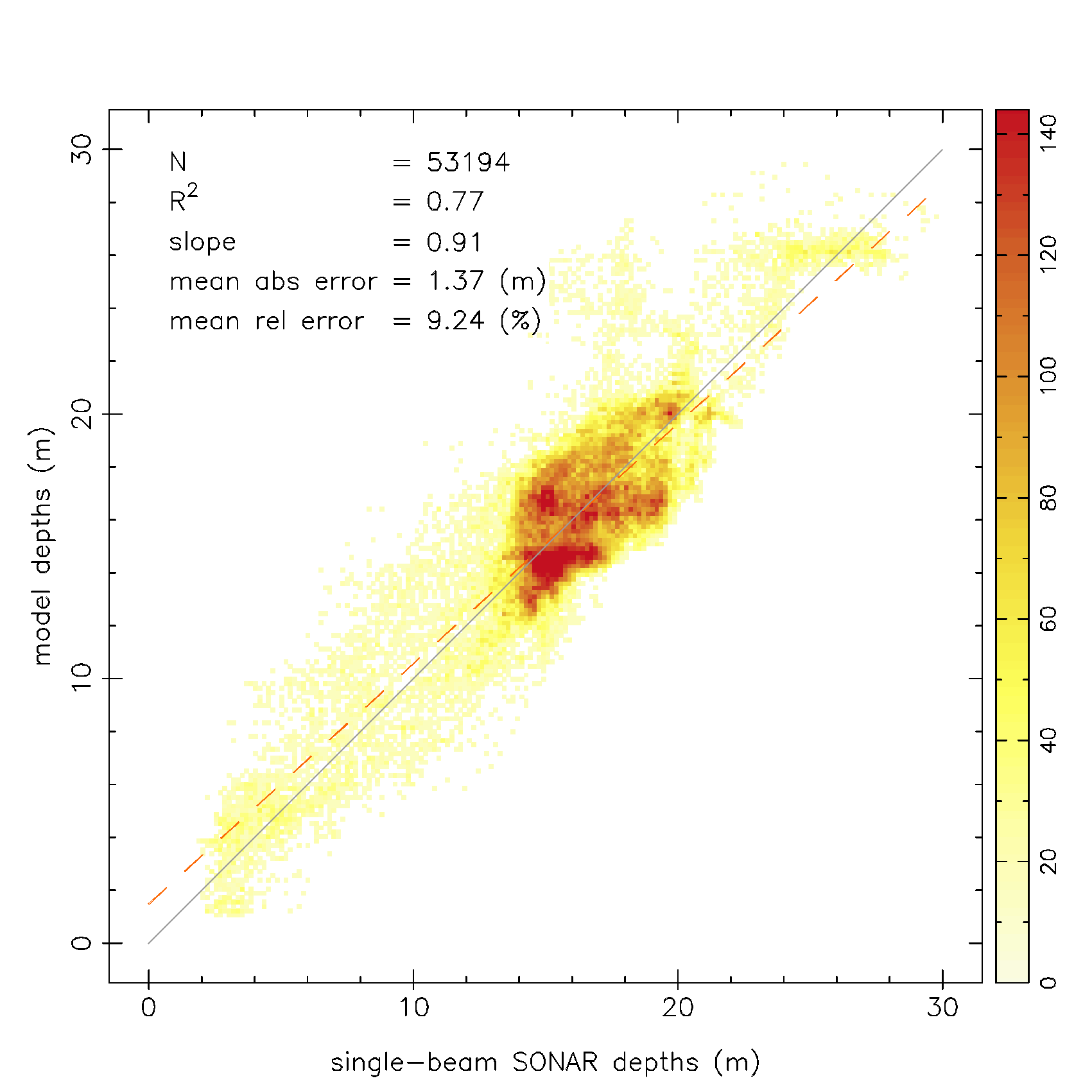}
\caption{A scatter (density) plot of the 2011 single-beam sonar bathymetry survey 
with the LANDSAT-derived bathymetry.}
\end{figure}
The regression analysis between the sonar survey and the LANDSAT-derived bathymetry 
shows good agreement, with $N = 53\,194$, $R^2 = 0.77$, the least squares line of 
best fit was $y = 0.91 x + 1.61$, the mean absolute error was 1.17 m, 
and the mean relative error was 7.52\%. Furthermore, below we give a summary of the 
absolute and relative errors. 

\begin{table}[H]
\centering
\caption{Summary of absolute and relative percentage errors.}
\begin{tabular}{ll|ll}
\begin{tabular}{r c c}
\hline
 9.94\% & within & 0.25 m \\
20.20\% & within & 0.50 m \\ 
31.59\% & within & 0.75 m \\
42.86\% & within & 1.00 m \\
62.58\% & within & 1.50 m \\
76.79\% & within & 2.00 m \\
\hline
\end{tabular}
&
\hspace{0.5cm}
&
\hspace{0.5cm}
&
\begin{tabular}{r c r}
\hline
12.51\% & within &  2.00 (rel \% error) \\
33.10\% & within &  5.00 (rel \% error) \\
64.89\% & within & 10.00 (rel \% error) \\
85.21\% & within & 15.00 (rel \% error) \\
94.62\% & within & 20.00 (rel \% error) \\
96.42\% & within & 25.00 (rel \% error) \\
\hline
\end{tabular}
\end{tabular}
\end{table}

The LANDSAT-derived bathymetry correlates well with the single-beam sonar 
data. While it is difficult to compare studies in different regions, the 
regression analysis suggests our model-derived bathymetry has similar 
performance to other physics-based, model-derived 
bathymetry \cite{ohlendorf2011}\cite{sagar2010}. We do not, however see 
LANDSAT 8 or Sentinel-2 data competing with hyperspectral data in terms 
of accuracy of bathymetric retrieval. \\

In the following graphic we display the key model-derived datasets and the model 
derived parameterisation of P, G and X. 

\begin{landscape}
\begin{figure}[H]
\centering
\includegraphics[height=0.8\textheight]{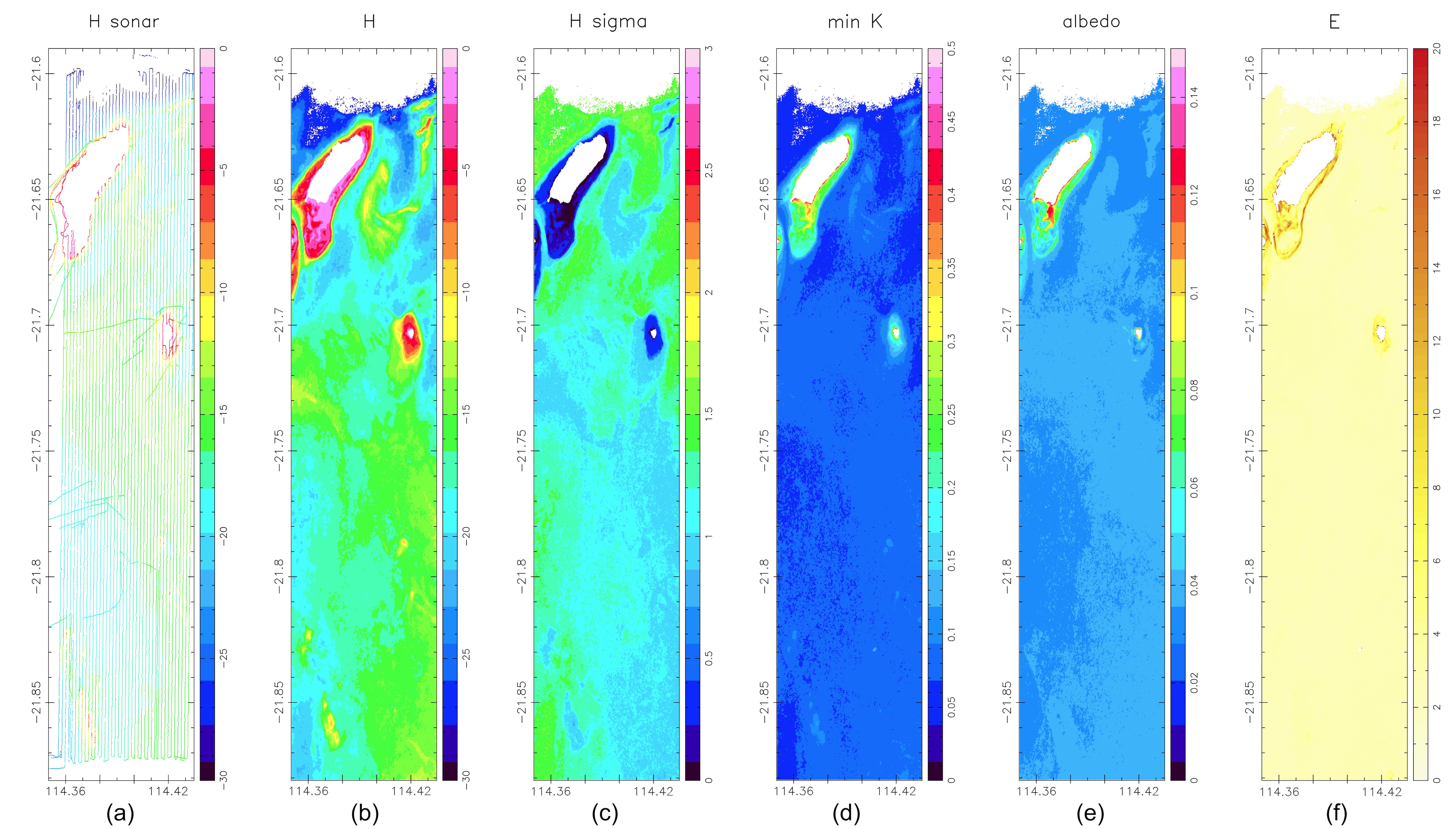}
\caption{A plot of \textbf{(a)} the sonar depths; \textbf{(b)} the model-derived depths, 
comprising the median of all 15 model iterations; \textbf{(c)} the one sigma 
depth error estimate; \textbf{(d)} the minimum attenuation coefficient, averaged over 
all 15 model iterations; \textbf{(e)} the model-derived bottom albedo (the parameter 
$B$ from (\ref{bottom_formulation})); \textbf{(f)} the model error, 
$E_{\text{photic}}$, averaged over all 15 model iterations.}
\end{figure}
\end{landscape}

\begin{landscape}
\begin{figure}[H]
\centering
\includegraphics[height=0.8\textheight]{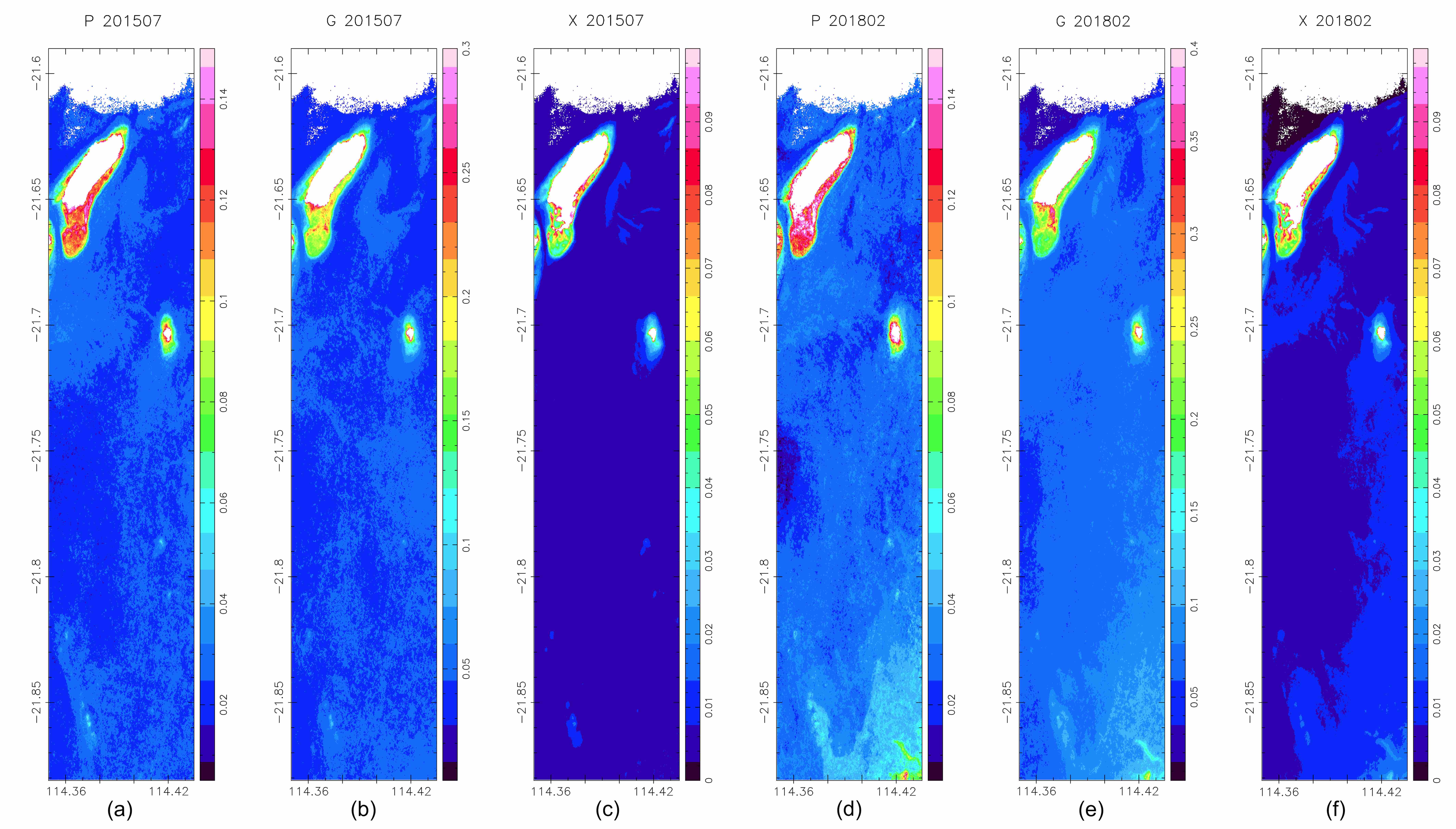}
\caption{A plot of P, G and X for scenes \texttt{LC81150752015210LGN02} and \texttt{LC81150752018058LGN00}. In
each of these plots we have taken the median of the 15 model iterations.}
\end{figure}
\end{landscape}

\begin{landscape}
\begin{figure}[H]
\centering
\includegraphics[height=0.8\textheight]{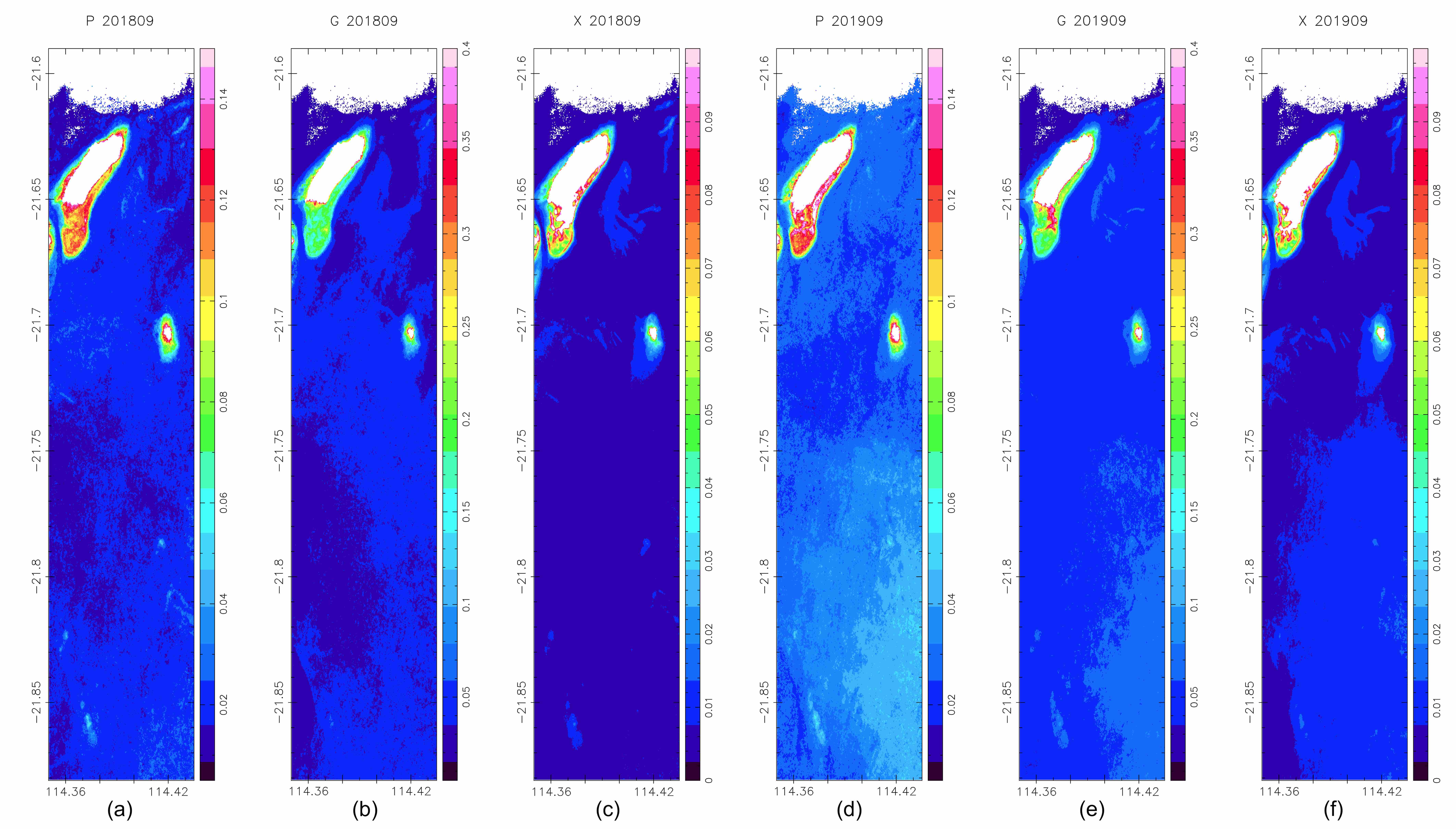}
\caption{A plot of P, G and X for scenes \texttt{LC81150752018266LGN00} and \texttt{LC81150752019253LGN00}. In
each of these plots we have taken the median of the 15 model iterations.}
\end{figure}
\end{landscape}

It is of interest to see the increase in $E_{\text{photic}}$ in a band to the south of 
North Murion Island (plot (f) above). This coincides with a sharp drop-off in depth 
and consequently is can be explained by the model trying to minimise both the RRS error 
and the depth continuity error (\ref{error_H}). As the RRS error is significantly higher 
weighted than the depth continuity error, the model will favour the minimisation of the 
RRS error in such cases. Otherwise, the average error is well below 5\%, which is 
expected. \\

The range of H-sigma is reasonable, as this estimate does not include uncertainties 
introduced in the estimation of the bottom spectra, atmospheric correction, 
and core assumptions within the original model of Lee et al. \\

When interpreting the range of the mean, minimum attenuation coefficient it must be 
considered that the 4 scenes were selected amongst over 100 scenes for their 
minimal water turbidity. \\

One region where upon visual inspection the model-derived bathymetry performs poorly 
is directly east of North Murion Island. Below we have plotted the sonar data over 
the model-derived bathymetry. This may be due to a shift in the seabed in the 9 years 
since the sonar survey was conducted (unlikely) or due to poorly estimating the bottom 
reflectance in this area (likely). 

\begin{figure}[H]
\centering
\includegraphics[width=0.66\textwidth]{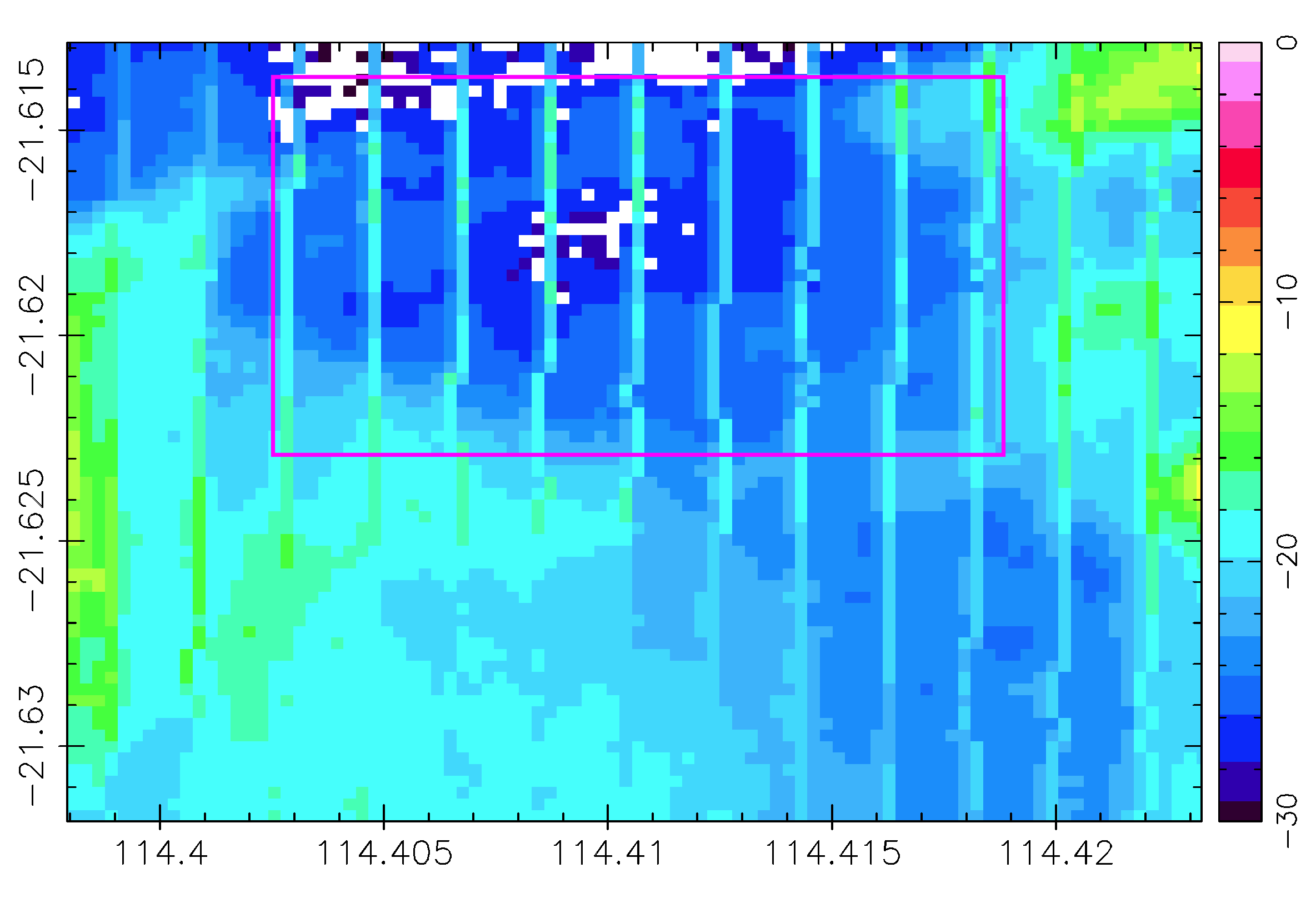}
\caption{A region east of North Murion Island (highlighted with a pink rectangle) where 
the model-derived bathymetry performs poorly. The single-bean sonar data is displayed  
over the model-derived bathymetry.}
\end{figure}

By way of comparison, we modelled each scene individually with the spatial and temporal 
extensions removed ($N_r = 1$, $N_s = 1$). We also removed the initial depth estimates 
and the lookup table. The resulting model-derived bathymetry clearly shows instabilities 
in the optimisation scheme with so few data points to perform the fit. 

\begin{figure}[H]
\centering
\includegraphics[width=0.66\textwidth]{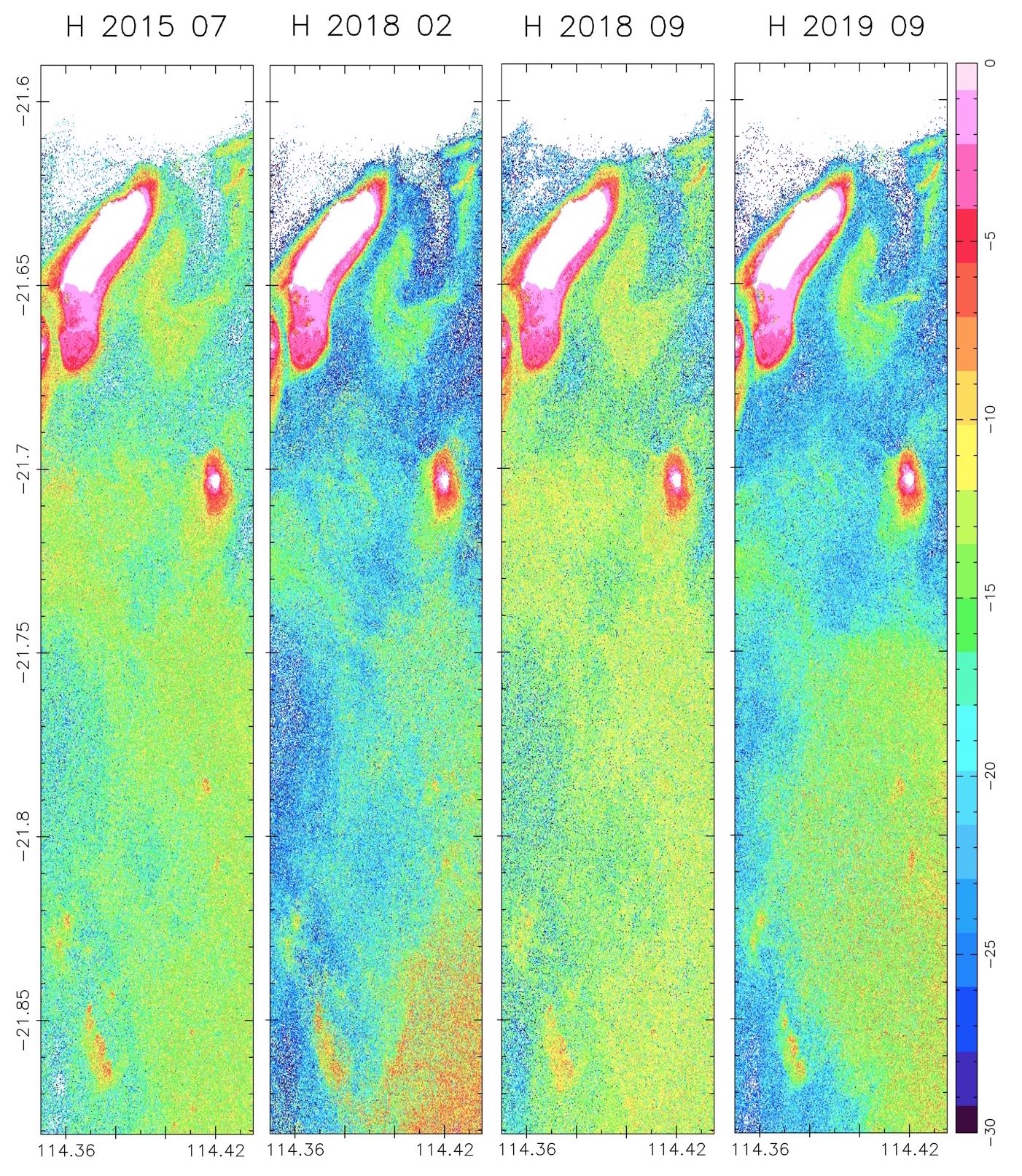}
\caption{Plot of the model-derived bathymetry with the parameterisation 
$N_r = 1$, $N_s = 1$, $N_b = 3$, and the initial depth estimates and lookup 
table removed from the model.}
\end{figure}


\section{Conclusions \& Future Work}

In this paper we have shown that multispectral imagery can be used successfully within a 
physics-based. We have made temporal and spatial extensions to the model of Lee et al 
\cite{lee1998}\cite{lee1999}\cite{lee2001}, which increase the stability and accuracy 
of depth retrievals when using low spectral resolution satellites like LANDSAT 8 or 
Sentinel-2. \\

Further work is underway to sort the spectra into clusters, where in each cluster only a 
small fraction of the spectra require modelling due to their spectral similarity. 

\bibliographystyle{abbrv}

\appendix
\section{Appendix}
From the model of Lee for $a_{\phi}(\lambda) = \left(a_0(\lambda) + a_1(\lambda)\log(P)\right)P$, 
we derived (using least-squares, non-linear optimisation) $a_0$ and $a_1$ using a dataset 
collected by the CSIRO in 2003--2005 near Bunbury, Western Australia \cite{csiro2005}. This 
dataset comprised 71 measurements and collected in depths ranging from 1.5 to 20 m. 

\begin{table}[H]
\centering
\caption{440nm-normalised $a_0$, $a_1$ derived from the Bunbury, Western Australia 
field measurements.}
\begin{tabular}{ccc|ccc|ccc}
 $\lambda$ & $a_0$ & $a_1$ & $\lambda$ & $a_0$ & $a_1$ & $\lambda$ & $a_0$ & $a_1$ \\
 \hline 
 400 & 0.69322 & 0.01035 & 520 & 0.32875 & 0.00617 & 640 & 0.24139 & 0.02122 \\
 405 & 0.80506 & 0.01868 & 525 & 0.30033 & 0.00753 & 645 & 0.25922 & 0.02508 \\
 410 & 0.89891 & 0.02278 & 530 & 0.27633 & 0.00874 & 650 & 0.26483 & 0.02589 \\
 415 & 0.96392 & 0.02497 & 535 & 0.25874 & 0.01065 & 655 & 0.27135 & 0.02374 \\
 420 & 0.99268 & 0.02371 & 540 & 0.23621 & 0.01005 & 660 & 0.31442 & 0.02326 \\
 425 & 1.00392 & 0.01841 & 545 & 0.21342 & 0.00897 & 665 & 0.40322 & 0.02714 \\
 430 & 1.02963 & 0.01381 & 550 & 0.19724 & 0.00975 & 670 & 0.49153 & 0.03177 \\
 435 & 1.03967 & 0.00750 & 555 & 0.18247 & 0.01048 & 675 & 0.52301 & 0.03344 \\
 440 & 1.0 & 0.0 & 560 & 0.16819 & 0.01044 & 680 & 0.46490 & 0.02943 \\
 445 & 0.90067 & -0.01143 & 565 & 0.15781 & 0.01023 & 685 & 0.33078 & 0.01968 \\
 450 & 0.79228 & -0.02292 & 570 & 0.15495 & 0.01076 & 690 & 0.19484 & 0.01007 \\
 455 & 0.74203 & -0.02655 & 575 & 0.15478 & 0.01080 & 695 & 0.11332 & 0.00577 \\
 460 & 0.74870 & -0.02273 & 580 & 0.15795 & 0.01102 & 700 & 0.07804 & 0.00588 \\
 465 & 0.76773 & -0.01590 & 585 & 0.16251 & 0.01104 & 705 & 0.05617 & 0.00477 \\
 470 & 0.77611 & -0.00746 & 590 & 0.16427 & 0.01054 & 710 & 0.04426 & 0.00401 \\
 475 & 0.76177 & -0.00132 & 595 & 0.16247 & 0.01027 & 715 & 0.03844 & 0.00414 \\
 480 & 0.72663 & -0.00007 & 600 & 0.16094 & 0.01062 & 720 & 0.03209 & 0.00361 \\
 485 & 0.68161 & -0.00094 & 605 & 0.16188 & 0.01104 & 725 & 0.02705 & 0.00333 \\
 490 & 0.63211 & -0.00109 & 610 & 0.16489 & 0.01075 & 730 & 0.02090 & 0.00253 \\
 495 & 0.57497 & -0.00056 & 615 & 0.17238 & 0.01088 & 735 & 0.02198 & 0.00528 \\
 500 & 0.51537 & 0.00073 & 620 & 0.17878 & 0.01073 & 740 & 0.01671 & 0.00383 \\
 505 & 0.45850 & 0.00244 & 625 & 0.18479 & 0.01090 & 745 & 0.00866 & 0.00191 \\
 510 & 0.40764 & 0.00391 & 630 & 0.19970 & 0.01329 & 750 & 0.01262 & 0.00288 \\
 515 & 0.36526 & 0.00529 & 635 & 0.21805 & 0.01636 &  &  &  \\
\end{tabular}
\end{table}

Below we compare our derived $a_0$, $a_1$ with the original parameterisation of 
Lee \cite{lee1994} and Lee et al \cite{lee1998} against 12 of the collected samples.

\begin{figure}[H]
\centering
\includegraphics[width=0.95\textwidth]{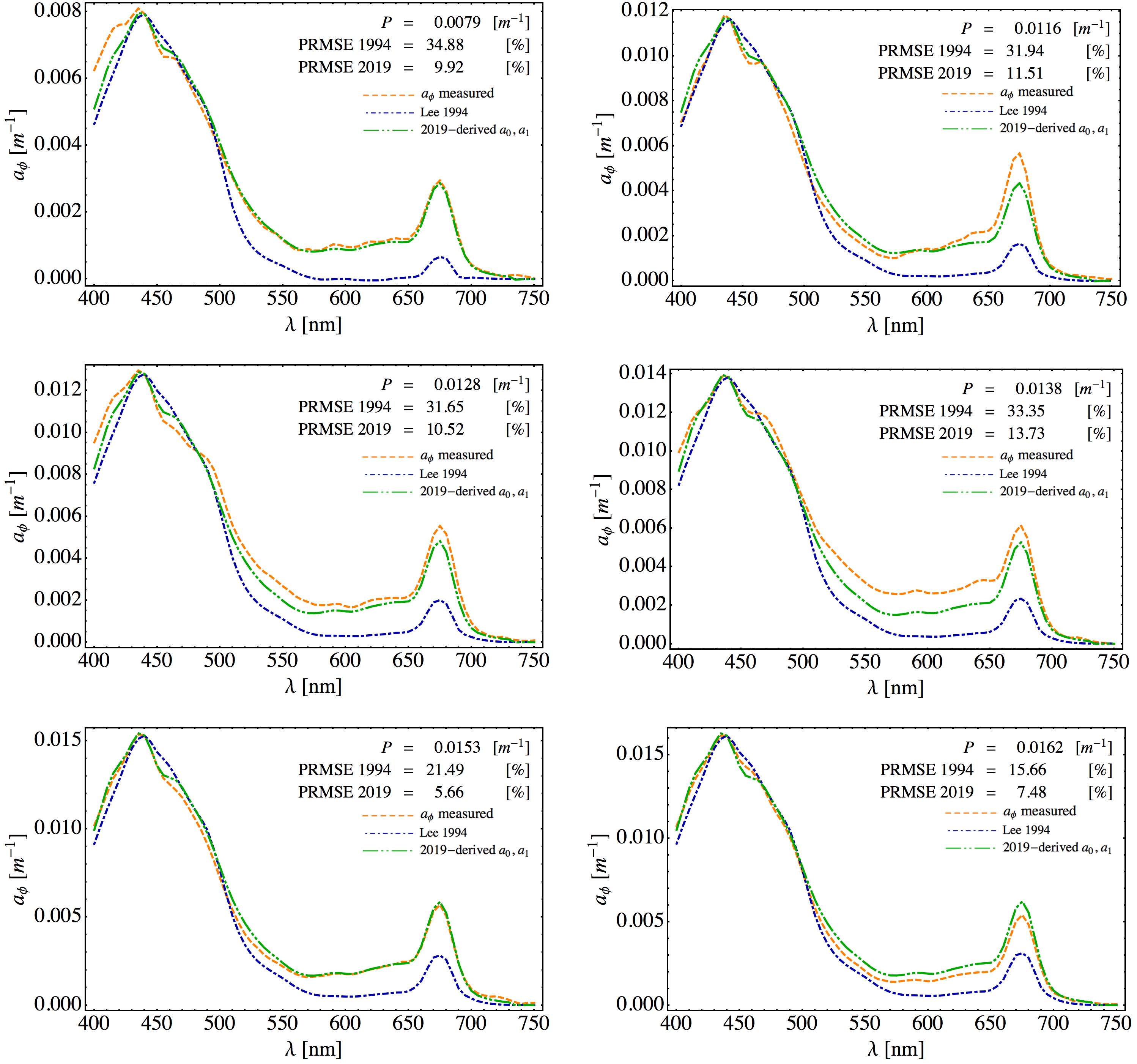}
\caption{A plot comparing the measured and modelled $a_\phi$ for the original 
parameterisation of Lee and the Bunbury-derived $a_0$, $a_1$.}
\end{figure}

\begin{figure}[H]
\centering
\includegraphics[width=0.95\textwidth]{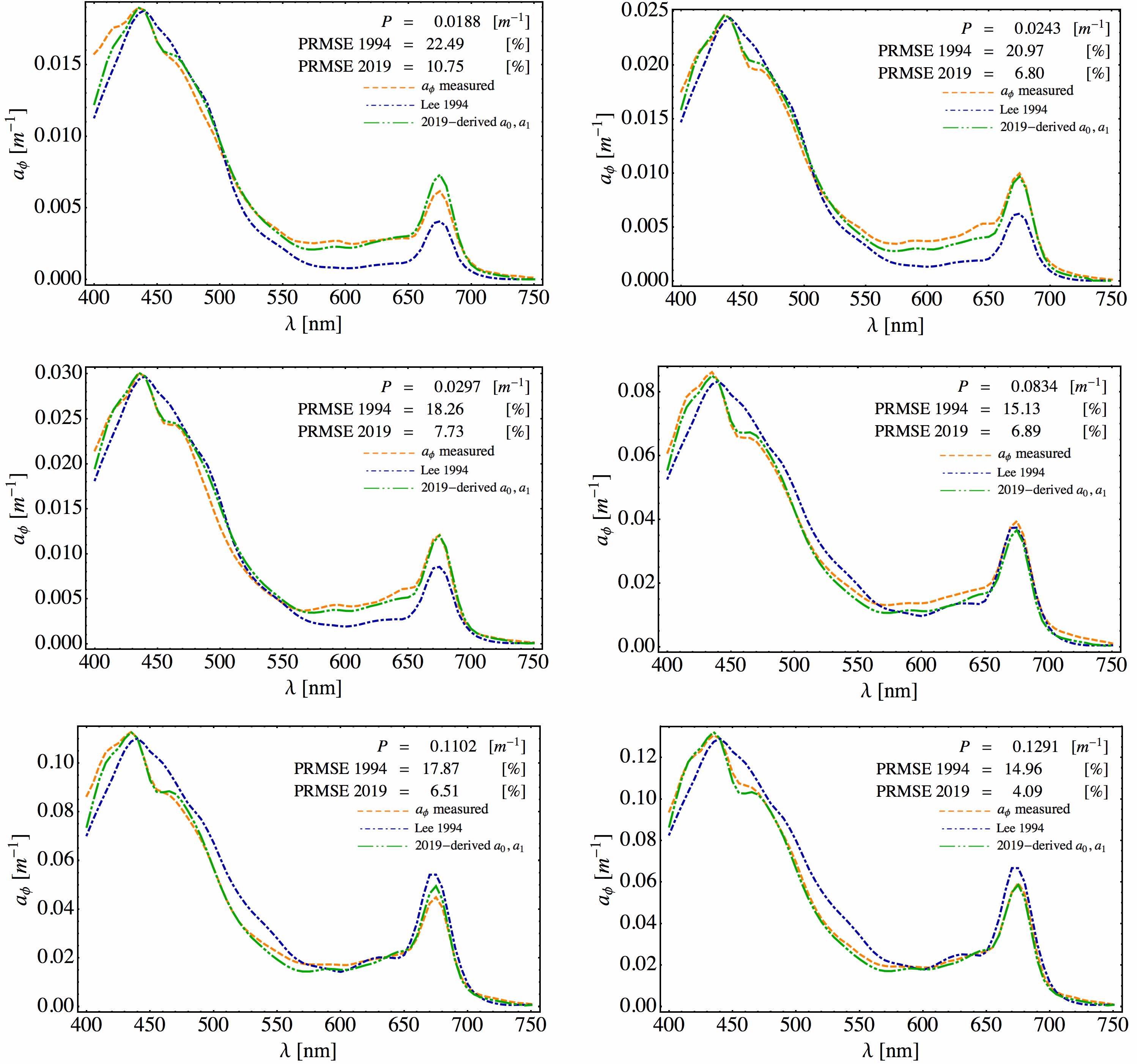}
\caption{A plot comparing the measured and modelled $a_\phi$ for the original 
parameterisation of Lee and the Bunbury-derived $a_0$, $a_1$.}
\end{figure}

\end{document}